\documentclass[fleqn,usenatbib]{mnras}

\usepackage{mathptmx}
\usepackage[T1]{fontenc}
\usepackage{ae,aecompl}
\usepackage{graphicx} 
\usepackage{amsmath} 
\usepackage{amssymb} 
\usepackage{verbatim} 
\usepackage[english]{babel}
\usepackage[dvipsnames]{xcolor}
\usepackage{natbib}
\usepackage{txfonts}
\usepackage{multirow}
\usepackage{pifont}

\newcommand{\ramses}{\textsc{ramses}}

\newcommand{\sub}[1]{_{\mathrm{#1}}}

\title[ICM loss in GCs]{The loss of the intra-cluster medium in globular clusters}

\author[W. Chantereau et al.]{
W. Chantereau,$^{1}$\thanks{E-mail: w.chantereau@ljmu.ac.uk}
P. Biernacki,$^{2,3}$
M. Martig,$^{1}$
N. Bastian,$^{1}$
M. Salaris,$^{1}$
R. Teyssier$^{3}$
\\
$^{1}$Astrophysics Research Institute, Liverpool John Moores University, 146 Brownlow Hill, Liverpool L3 5RF, UK \\
$^{2}$Kavli Institute for Cosmology \& Institute of Astronomy, University of Cambridge, Madingley Road, Cambridge CB3 0HA, UK \\
$^{3}$Center for Theoretical Astrophysics and Cosmology, Institute for Computational Science,\\     \quad University of Zurich, Winterthurerstrasse 190, 8057 Zurich, Switzerland \\
}

\date{Accepted XXX. Received YYY; in original form ZZZ}

\pubyear{2019}

\begin{document}

\maketitle

\begin{abstract}
Stars in globular clusters (GCs) lose a non negligible amount of mass during their post-main sequence evolution. This material is then expected to build up a substantial intra-cluster medium (ICM) within the GC. However, the observed gas content in GCs is a couple of orders of magnitude below these expectations. Here we  follow the evolution of this stellar wind material through hydrodynamical simulations to attempt to reconcile theoretical predictions with observations. 
We test different mechanisms proposed in the literature to clear out the gas such as ram-pressure stripping by the motion of the GC in the Galactic halo medium and ionisation by UV sources. We use the code \ramses~to run 3D hydrodynamical simulations to study for the first time the ICM evolution within discretised multi-mass GC models including stellar winds and full radiative transfer. We find that the inclusion of both ram-pressure and ionisation is mandatory to explain why only a very low amount of ionised gas is observed in the core of GCs. The same mechanisms operating in ancient GCs that clear the gas could also be efficient at younger ages, meaning that young GCs would not be able to retain gas and form multiple generations of stars as assumed in many models to explain "multiple populations". However, this rapid clearing of gas is consistent with observations of young massive clusters.
\end{abstract}


\section{Introduction}\label{introduction}

Globular clusters (GCs) in the Galaxy tend to be old (9-13~Gyr) and massive (from $\sim$10$^4$ up to $\sim$10$^6$~M$_\odot$). They are located from a few kpc up to $\sim$100~kpc from the Galactic centre and they orbit the Galaxy in a few hundreds Myr \citep[e.g.,][]{Odenkirchen97}. GCs contain hundreds to thousands of stars along their red giant branches which display significant mass-loss (also along the asymptotic giant branch to some extent). This mass is lost through mostly\footnote{Note however that fast outflows can also be found, but it represents a minority \citep{Dupree09}.} slow winds \citep[$\sim$10-20 km~s$^{-1}$, e.g.,][]{Mauas06,McDonald07,Meszaros09}, well below the central escape velocities of GCs \citep[of the order of a few km~s$^{-1}$ up to $\sim$90~km~$^{-1}$, e.g.,][]{Baumgardt18}. The mass lost by the whole stellar population of a massive GC ($\sim$10$^6$~M$_\odot$) can be approximated by $\sim10^{-6}$~M$_\odot$ yr$^{-1}$ \citep[e.g.,][]{McDonald15}. GCs cross the Galactic disc over time-scales of a few hundreds Myr, where they are subject to ram-pressure which is efficient at stripping all their ICM \citep{Roberts88,Tayler75}. Thus stellar winds contribute to build up a non-negligible, observable ICM of a few hundreds of M$_\odot$ between two Galactic disc-crossings. 

However, at odds with these theoretical considerations, to date there has been only a single certain detection of neutral intra-cluster medium (ICM) in GCs: M15 (NGC~7078) displays 0.3~M$_\odot$ of neutral hydrogen HI and a very low amount of dust (9$\times10^{-4}$~M$_\odot$) in its core \citep{Evans03,Boyer06,vanLoon06}. In addition, \cite{Freire01} inferred from the plasma mass density detected in 47~Tuc (NGC~104) that it should contain a total mass of $\sim$0.1~M$_\odot$ of gas within its inner 2.5~pc. Recently, \cite{Abbate18} calculate a total mass of gas in the inner 1 pc of 47~Tuc to be 0.023$\pm$0.005~M$_\odot$. Other searches for HI, CO, and H$\alpha$ in GCs all lead to only observational upper limits often lower than 1~M$_\odot$ and as low as  10$^{-5}$~M$_\odot$ \citep[e.g.,][and references therein]{Roberts88,Smith90,vanLoon06,Boyer08,Matsunaga08,Barmby09,vanLoon09}. 

Therefore to explain the negligible amount of ICM observed in GCs, a mechanism to get rid of the gas with a time-scale much lower than a few hundreds Myr should be at work. Several processes have been proposed in the literature to explain this missing ICM. \cite{Coleman77} suggested that flaring activity from M-dwarf stars would help to clear the GC from its ICM, however this process is subject to several uncertainties (e.g. mass-loss rates). \cite{Vandenberg78} suggested hot horizontal-branch stars to heat the ICM thanks to their UV-radiation, but the main drawback is that these stars are not present in all GCs. Classical novae have been proposed as an internal clearing mechanism, however novae are effective at clearing the ICM only in low-mass GCs \citep[$\leq 10^5$~M$_\odot$,][]{Scott78,Moore11}. \cite{Spergel91} and \cite{Yokoo92} considered winds of pulsars and X-ray bursters, respectively, as gas removal mechanisms. However, if they are present in the stellar cluster, they are not effective enough to heat the ICM \citep{McDonald15}. \cite{Thoul02} briefly mentioned that the mass lost through winds could be recycled by accretion onto low-mass stars. \cite{Umbreit08} invoked stellar collisions to remove the gas from M15 on time-scales of $\sim$10$^6$~yr. Finally \cite{Dupree09} suggested that stellar winds of RGB stars were fast enough to leave the cluster potential. However, most of the studies in the literature found that velocities of low-mass giant winds are rather low and range between a few and $\sim$20 km~s$^{-1}$ \citep[e.g.,][]{Netzer93,Mauas06,McDonald07,Meszaros09,Groenewegen14}.

\cite{Frank76} and \cite{Priestley11} investigated the ram-pressure stripping of the ICM as the GC moves through the Galactic halo. However, \cite{Priestley11} has shown with a hydrodynamical study\footnote{They did not model full radiative transfer and took into account a discrete, orbiting stellar population in their simulations at low resolution only as exploratory stages.} that this mechanism is sufficient to limit the amount of ICM to levels similar to what it is observed only within intermediate-mass GCs ($\sim 10^5$~M$_\odot$). Thus an additional process is needed to explain observations in massive GCs ($\sim 10^6$~M$_\odot$). Recently, \cite{McDonald15} investigated the UV radiation from cooling WDs and hot post-AGB stars in stellar clusters and they showed with a 1D model that the ICM of every GCs should be ionised\footnote{Note that it has also been shown with 3D hydrodynamical models that photoionisation have a great impact on stellar formation in young star clusters \citep{Geen15,Gavagnin17}.}, and in turn, the ICM would expand beyond the GC's tidal radius and would escape its gravitational potential. However, in this case full hydrodynamic models would be mandatory to take into account different physical mechanisms at work and better describe e.g. the interplay between the ICM and the Galactic halo.  

In this paper we follow the different suggestions of \cite{Priestley11} and \cite{McDonald15} and test the effects of ram-pressure stripping by the motion of the stellar cluster in the Galactic halo medium and the inclusion of internal ionising sources on the ICM evolution in GCs. We investigate these two mechanisms in  intermediate-mass and massive GCs to propose a general solution to the lack of gas in GCs and explain all the observations. We use full 3D hydrodynamical simulations with the adaptive mesh refinement (AMR) code \ramses~to follow for the first time the ICM evolution within discretised multi-mass GC models\footnote{The stellar population of the cluster is represented by modelling each individual star with different masses and orbiting in the cluster potential.} taking into account stellar winds, ionising radiation, radiative heating and radiative pressure. We present the code and the setup of the two sets of simulations for typical intermediate-mass (Simu1, $\sim 10^5$~M$_\odot$) and a massive GC such as 47~Tuc (Simu2, $\sim 10^6$~M$_\odot$) in Sect.~\ref{ramses}. We discuss the different effects of the ram-pressure stripping and ionisation mechanisms on the ICM evolution in these GCs in Sect.~\ref{results}. Finally we summarise the results and discuss their implications in Sect.~\ref{discussion}.

\section{Simulation setup}\label{ramses}

We have performed 3D simulations using the AMR code \ramses{} \citep{Teyssier2002}, which uses the second-order, unsplit Godunov scheme to solve the Euler equations. The dynamical evolution of the stars employs the Adaptive Particle-Mesh solver with cloud-in-cell interpolation. We do not allow new stars to form and we do not include any supernova feedback modelling, as we do not expect any star formation or stars to explode during the time of evolution of our simulations.  We do not model any external potential that would come from the halo/galaxy. The cooling of gas follows the prescription given by \citet{Sutherland1993} for radiative cooling of gas for H, He and metal lines if the gas is hotter than $10^4\,\mathrm{K}$ and from metal fine-structure cooling processes at lower temperatures. We advect the metallicity in the form of a passive scalar and we follow it in each cell. A temperature floor is introduced at 50~K. It prevents the gas from artificial fragmentation by matching the corresponding Jeans length to our adopted resolution. The mesh refinement strategy we have adopted for all our simulations is a quasi-Lagrangian approach, where cells are refined once their mass exceed $1.25\times10^{-4}\,M\sub{\sun}$. The maximum spatial resolution achieved by our simulations is 0.098-0.146\,pc for a box length of 100-150~pc for the first and the second set of simulations, respectively\footnote{The box radius is larger than the typical tidal radius of intermediate-mass GCs (first set of simulations) and than massive GCs (second set of simulations) to not lose any information \citep[e.g., r$_t \sim52.5$~pc for 47~Tuc,][]{deBoer19}}. The minimum and maximum levels of refinement are 8 and 10, respectively. We model at high resolution for each simulation individual stars orbiting in the cluster potential. We use as a boundary condition free outflow. In simulations where the full radiative transfer is used \citep[\textsc{ramses-rt},][]{Rosdahl2013,Rosdahl2015a} we use the non-equilibrium cooling. As in \cite{Rosdahl2015b} we group the photons into five bins, with the same energies, which we list in \autoref{tab:rt}. The energy fraction per photon group follows the average spectral energy distribution (SED) of a hot post-AGB star (cf. next section and Sect.~\ref{Simu2}).

\begin{table*}
	\begin{center}
	\begin{tabular}{lrrrrrrr}
	\hline
	Photon group & $\epsilon_0$ [eV]& $\epsilon_1$ [eV] & $\sigma_{\mathrm{H}_{\mathrm{I}}}$ [cm$^2$] & $\sigma_{\mathrm{He}_{\mathrm{I}}}$ [cm$^2$] & $\sigma_{\mathrm{He}_{\mathrm{II}}}$ [cm$^2$] & $\tilde{\kappa}$ [$\mathrm{cm}^2\,\mathrm{g}^{-1}$] & $f\sub{\gamma,i}$\\
	\hline
	IR								& 0.10	& 1.00	& 0 & 0 & 0 & 10 		& 0.2289\\
	Opt								& 1.00	& 13.60	& 0 & 0 & 0 & 1000		& 0.3759\\
	UV$_{\mathrm{H_{\mathrm{I}}}}$		& 13.60	& 24.59	& $3.3\times10^{-18}$ & 0 & 0 & 1000	& 0.0829\\
	UV$_{\mathrm{He_{\mathrm{I}}}}$		& 24.59	& 54.42	& $6.3\times10^{-19}$ & $4.8\times10^{-18}$ & 0 & 1000	& 0.0695\\
	UV$_{\mathrm{He_{\mathrm{II}}}}$	& 54.42	& $\infty$& $9.9\times10^{-20}$ & $1.4\times10^{-19}$ & $1.3\times10^{-18}$ & 1000	& 0.1243\\
	\hline
	\end{tabular}
	\caption{Properties of the photon groups used in the radiative transfer simulations. The energy intervals defined by the groups are indicated by $\epsilon_0$ and $\epsilon_1$. $\sigma_{\mathrm{H}_{\mathrm{I}}}$, $\sigma_{\mathrm{He}_{\mathrm{I}}}$ and $\sigma_{\mathrm{He}_{\mathrm{II}}}$ denote the cross-sections for ionisation of hydrogen and helium, respectively. $\tilde{\kappa}$ is the dust opacity. $f\sub{\gamma,i}$ is the reduced flux of the radiation of the group $i$  (describing the directionality of the radiation).}\label{tab:rt}
	\end{center}
\end{table*}

\subsection*{Initial conditions}

\begin{table*}
\centering
\begin{tabular}{ c | c c c c c c c}
	\hline 
	ID & M$_{cluster}$ (M$_\odot$) & V$_\mathrm{cluster}$ (km~s$^{-1}$) & $\dot{M}_{*}$ (M$_\odot yr^{-1}$) & T$_{winds}$ (K) & T$_{halo}$ (K) & $\rho_{halo}$ (cm$^{-3}$) & UV flux (s$^{-1}$) \\ \hline
    Simu1A & 10$^{5}$ & 0 & 3.2 $\times 10^{-7}$ & 6000 & 10$^{5.5}$ & 6 $\times 10^{-4}$ & 0 \\  
    Simu1B & 10$^{5}$ & 200 & 3.2 $\times 10^{-7}$ & 6000 & 10$^{5.5}$ & 6 $\times 10^{-4}$ & 0 \\
    Simu1C & 10$^{5}$ & 200 & 3.2 $\times 10^{-7}$ & 6000 & 10$^{5.5}$ & 6 $\times 10^{-3}$ & 0 \\
    Simu1D & 10$^{5}$ & 200 & 3.2 $\times 10^{-7}$ & 4000 & 10$^{5.5}$ & 6 $\times 10^{-4}$ & 0 \\ \hline
    Simu2A & 10$^{6}$ & 200 & 2.8 $\times 10^{-6}$ & 6000 & 10$^{5.5}$ & 7 $\times 10^{-3}$ & 0 \\
    Simu2B & 10$^{6}$ & 200 & 2.8 $\times 10^{-6}$ & 6000 & 10$^{5.5}$ & 7 $\times 10^{-3}$ & 5.7 $\times 10^{48}$ \\
    Simu2C & 10$^{6}$ & 200 & 2.8 $\times 10^{-6}$ & 6000 & 10$^{5.5}$ & 1 $\times 10^{-4}$ & 0 \\
    Simu2D & 10$^{6}$ & 200 & 2.8 $\times 10^{-6}$ & 6000 & 10$^{5.5}$ & 1 $\times 10^{-4}$ & 5.7 $\times 10^{48}$ \\
    Simu2E & 10$^{6}$ & 200 & 2.8 $\times 10^{-6}$ & 6000 & 10$^{5.5}$ & 7 $\times 10^{-3}$ & 2.43 $\times 10^{47}$ \\ \hline
    \hline 
\end{tabular}
\caption[]{Main characteristics of the different simulations. M$_{cluster}$ is the mass of the stellar cluster (M$_\odot$); V$_\mathrm{cluster}$ the velocity of the cluster in the halo (km~s$^{-1}$); $\dot{M}_{*}$ the total cluster stellar mass-loss rate (M$_\odot yr^{-1}$); T$_{winds}$ the temperature of the stellar winds (K); T$_{halo}$ the temperature of the halo gas (K); $\rho_{halo}$ the halo gas density (cm$^{-3}$); the UV flux (s$^{-1}$, $\lambda < 912~\AA$, i.e. $\epsilon > 13.6$ eV).} \label{table:simulations}
\end{table*}

In this study we follow the suggestions of \cite{Priestley11} and \cite{McDonald15} and investigate the effects of the ram-pressure and ionisation mechanisms on typical intermediate-mass (Simu1, $10^5$~M$_\odot$) and a massive GC such as 47~Tuc (Simu2, $\sim 10^6$~M$_\odot$). We choose and present below initial conditions for massive GCs similar to what is used in \cite{McDonald15} and for typical intermediate-mass GCs similar to what is used in \cite{Priestley11}. \\

We reproduce the stellar structure of globular clusters with discretised multi-mass models with the \textsc{limepy} code \citep{Gieles15} which aims at describing the phase-space density of stars in tidally limited and mass-segregated star clusters. The initial, non-rotating stellar distribution follows a King profile with a central gravitational potential of 7 \citep[][]{King66}, similar to the value used in \cite{Priestley11} of 7.5. 

The properties of the two sets of simulations are presented in \autoref{table:simulations}. 

The intermediate-mass GC is a general case (we do not associate it to a particular GC), thus we choose that its stellar population follows a  mass function with a slope $\alpha$ of -2.35 down to $\sim$0.1~M$_\odot$ \citep{Salpeter55}. Since we chose 47~Tuc as the massive GC, the stellar population follows its global mass function with $\alpha = -0.53$ down to $\sim$0.1~M$_\odot$ \citep{Baumgardt18}. Thus we have more than $\sim$500,000 and $\sim$3,000,000 stars in our $10^5$ and $10^6$ M$_\odot$ GCs, respectively. This difference between the two mass functions has only a negligible impact on the stellar structure of our cluster and does not have any effect on our results.

The initial conditions of our simulations feature a gas grid which is filled with an uniform medium at $10^{5.5}$~K. The medium density in the first set of simulations is $6\times10^{-4}$~cm$^{-3}$ ($\sim$10$^{-27}$~g.cm$^{-3}$) in agreement with the properties of a hot Galactic halo medium \citep{Spitzer56} and similar to what is used in \cite{Priestley11}. The medium density in the second set of simulations is $7\times10^{-3}$~cm$^{-3}$ in agreement with the properties of the Galactic halo medium where the globular cluster 47~Tuc evolves \citep{Taylor93,McDonald15}. We fix the metallicity to an initial value of $Z = 0.002$, like in \cite{Priestley11}, for the gas and stars, we keep this value between the different sets of simulations since the metallicity of 47~Tuc is close ($Z\sim0.004$, depending on the alpha enhancement).

\subsection*{Ionising source}\label{ssec:photo}

We do not include any UV radiation while modelling the intermediate-mass GC in the first set of simulations. In the second set of simulations (Simu2, massive GCs such as 47~Tuc, $\sim$10$^6$~M$_\odot$), we follow the suggestion of \cite{McDonald15} to investigate the effect of UV radiation from post-AGB stars. We include ionising radiation from a source, (i.e. radiative heating and radiative pressure) in several simulations of this set (\textit{Simu2B, Simu2D} and \textit{Simu2E}). To take into account this source, we need to provide a spectral energy distribution (SED) with its properties. 

We took as a typical post-AGB star for a massive GC such as 47~Tuc a model with an initial main sequence mass of $0.93~M_\odot$ at [Fe/H] = -0.72 with [$\alpha$/Fe] = +0.4 which has an age of 12.4~Gyr at the RGB-tip and a current mass of $\sim0.59~M_\odot$ \citep[\textsc{starevol},][]{Lagarde12}. Then, to estimate the ionising photon rate of this source, we determine the properties of this star (temperature and luminosity) by averaging by the time spent in the region where a hot post-AGB star/cooling WD is supposed to fully ionise the ICM of 47~Tuc \citep[T$_{eff}>$14,000~K and L$>$1~L$_\odot$,][]{McDonald15}. This ionising source has a resulting very high mean temperature of $\sim$95,000~K and a luminosity of $\sim10^{3.57}~L_\odot$, it is then a hot post-AGB star which has still not entered the cooling sequence. We then create a SED of a black body with this effective temperature. It gives us an average of 5.7$\times 10^{48}$ ionising photons s$^{-1}$ ($\lambda < 912 \AA$). This value is an order of magnitude higher than the value from the model from \citealt{McDonald15} (2.43$\times 10^{47}$ s$^{-1}$), it allows us to maximise the continuous ionising photons rate and thus test the limit of this ionising mechanism. However, both ionising photons rates are already a few orders of magnitude higher than the rate needed to fully ionise the ICM of 47~Tuc \citep[1.6$\times 10^{44}$ photons s$^{-1}$,][]{McDonald15}. Thus, we would expect that the difference between these values has only a negligible effect on the results of the simulations. The source is then associated to one of the star particle in the simulation which displays a mass typical of a post-AGB in this GC. In our simulations it is initially located in the cube at the position X,Y,Z = -0.066,-0.04,-0.32 in~pc, thus it is near the centre of the GC.

\subsection*{Stellar winds}\label{ssec:winds}
Stellar winds are expected to produce radially symmetric outflows. This is true if the stellar rotation is minimal and if the star is static with respect to the ICM. Stellar winds are in general a complicated function of stellar mass and metallicity and  vary with the age of a star. The gas ejected by a star would be characterised by its mass, temperature and velocity. Keeping these parameters in mind we introduce a simple stellar wind model, which can mimic the true stellar wind in hydrodynamical simulations.

Ideally, the gas injected by the star into a grid should be spread spherically around the star. In case of the AMR simulations one would be required to find all nearby grid elements and inject the right mass, momentum and pressure taking into account their size and distance. Here we propose a simpler approach, which statistically gives the same result. At each time step we inject the aforementioned  quantities \emph{only} in the cell in which the star resides with a randomly generated velocity vector and a magnitude corresponding to the chosen wind speed. It is apparent, that enough repetitions of these injections would lead to spherical distribution of the gas around a star (whose mass is reduced by the amount of mass injected into a cell). To make it correct in the global frame of reference, we add the velocity of the parent star to the velocity of its injected gas. For simplicity, we do not include the time dependence on the wind injection rate, as our simulations were not run over periods in which the wind rate could change significantly.

Here we consider only the mass-loss of RGB stars, since the number of AGB stars is negligible compared to the number of RGB stars. To summarise, only star particles on the upper RGB (log$\left(\frac{L}{L_\odot}\right) \gtrsim$ 2.0, as mass-loss occurs near the very end of the RGB evolutionary stage) inject gas to the grid in which it resides with constant rate, velocity, temperature and metallicity. 

In the first set of simulations (Simu1, intermediate-mass GC), we adjust the mass-loss rate per star to get a total mass-loss rate ($\sim3.2\times10^{-7}~M_\odot yr^{-1}$) similar to what is used in \cite{Priestley11}. In the second set of simulations (Simu2, massive GC), we use an average mass-loss computed with the stellar evolution code \textsc{starevol} for a $0.93~M_\odot$ star at [Fe/H] = -0.72 with [$\alpha$/Fe] = +0.4 (12.4~Gyr at the RGB-tip). The mean value of the mass-loss rate along the RGB is 1.3$\times$10$^{-9}~M_\odot$yr$^{-1}$ in our case \citep[mass-loss parameter $\eta$ = 0.5,][]{Reimers75}, and the total mass-loss rate of the stellar cluster is then $\sim$2.8$\times$10$^{-6}~M_\odot$yr$^{-1}$, similar to what is used in \cite{McDonald15}. We choose a typical value for the wind velocity of 20\,km~s$^{-1}$, the temperature of the wind is chosen to be the effective temperature of the progenitor. We choose 6,000~K which is an upper limit of the stellar T$_{eff}$ of RGB stars in metal-poor clusters (we will also test 4,000~K which is typical of stars of the upper RGB close to the tip in more metal rich globular clusters, Simu1D, see Table~\ref{table:simulations}). Finally each chosen star injects gas with $Z = 0.002 $, metallicity used for the simulations in \cite{Priestley11} and close to the metallicity of 47~Tuc ($Z\sim0.003$, depending on the alpha enhancement).  

\section{RESULTS}\label{results}

\subsection{Intermediate-mass GC - Simu1}

\begin{figure*}
    \centering
    \includegraphics[width=0.45\textwidth]{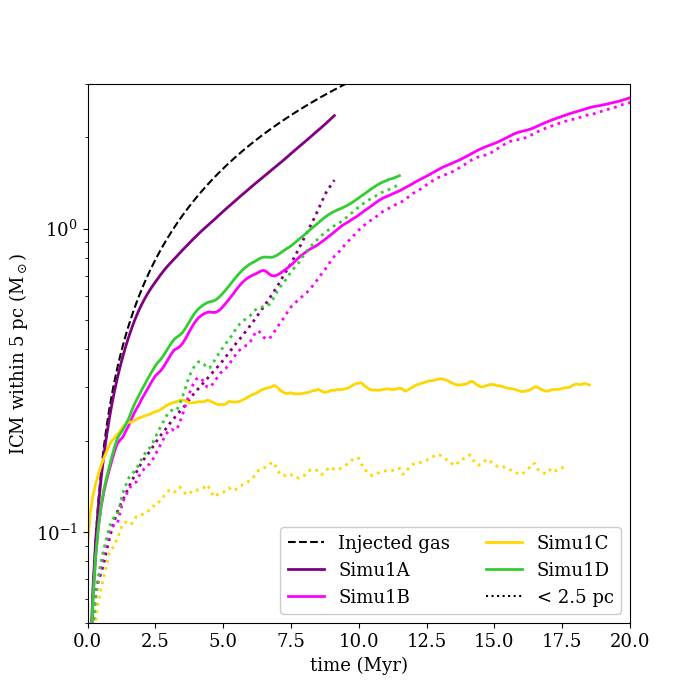}  
    \includegraphics[width=0.45\textwidth]{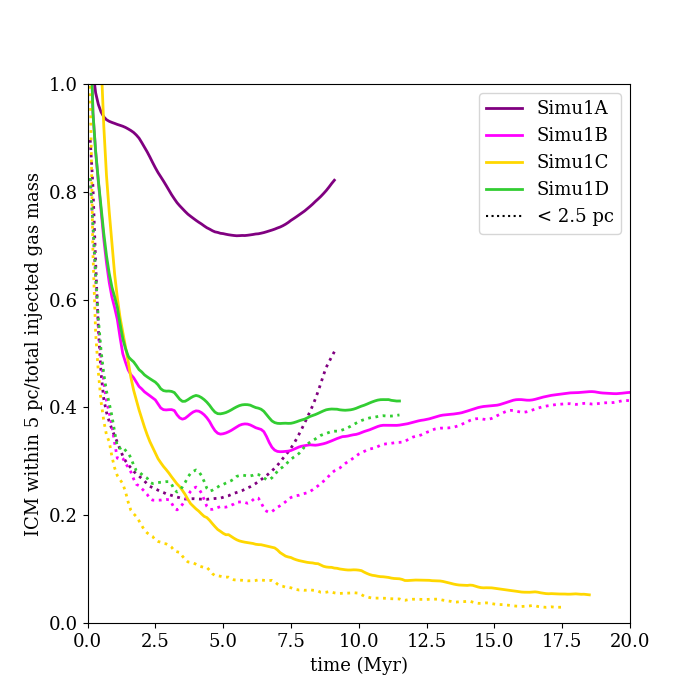} 
    \caption{\textit{Left:} ICM mass within 5~pc and the central part of 2.5~pc (solid and dotted lines, respectively) for the first set of simulations as a function of time. \textit{Right:} ratio of the ICM mass within 5~pc over the total mass injected by stellar winds for the first set of simulations as a function of time.}
    \label{Figure:mass}
\end{figure*}

\begin{figure*}
    \centering
    \includegraphics[width=0.49\textwidth]{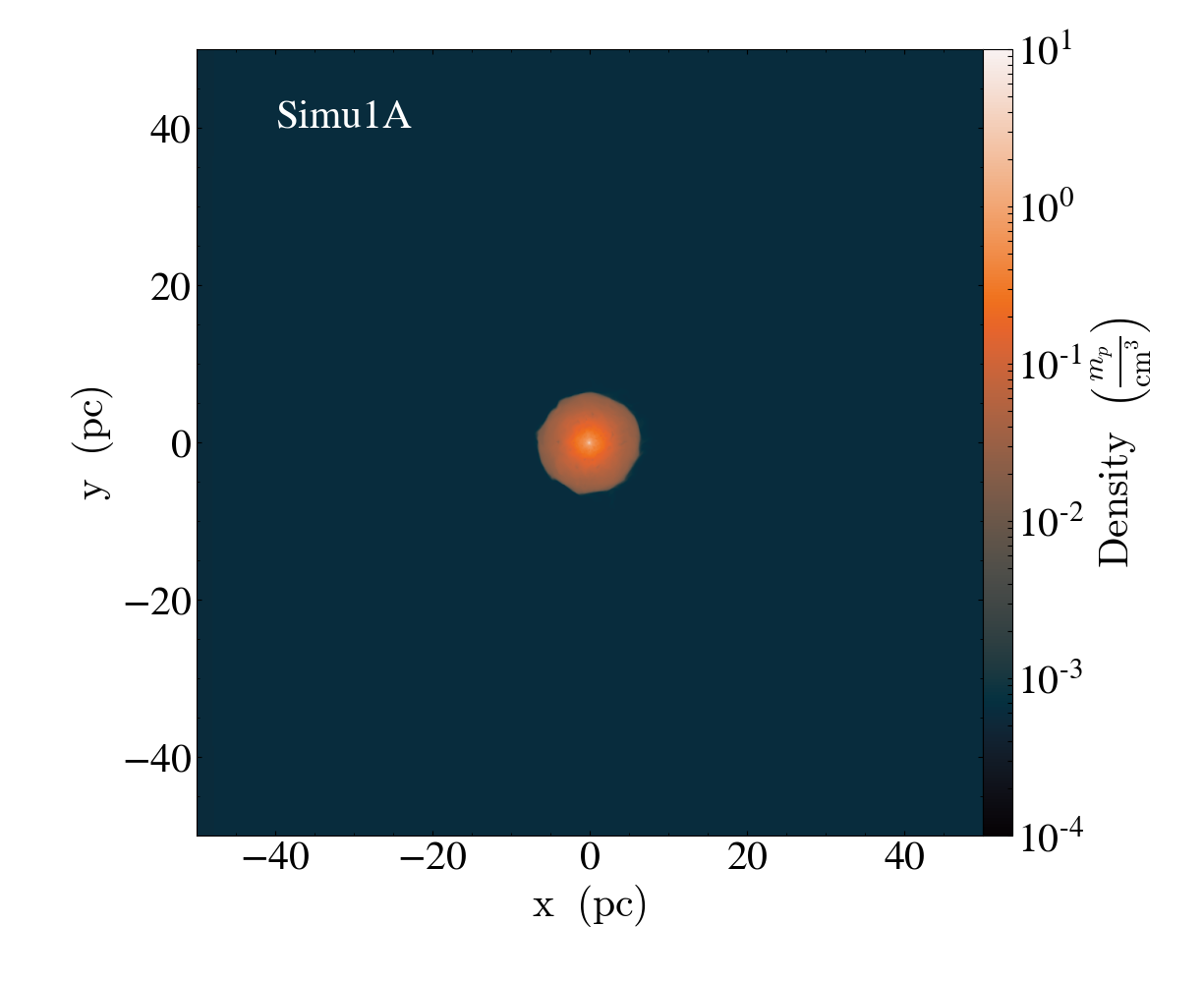}  
    \includegraphics[width=0.49\textwidth]{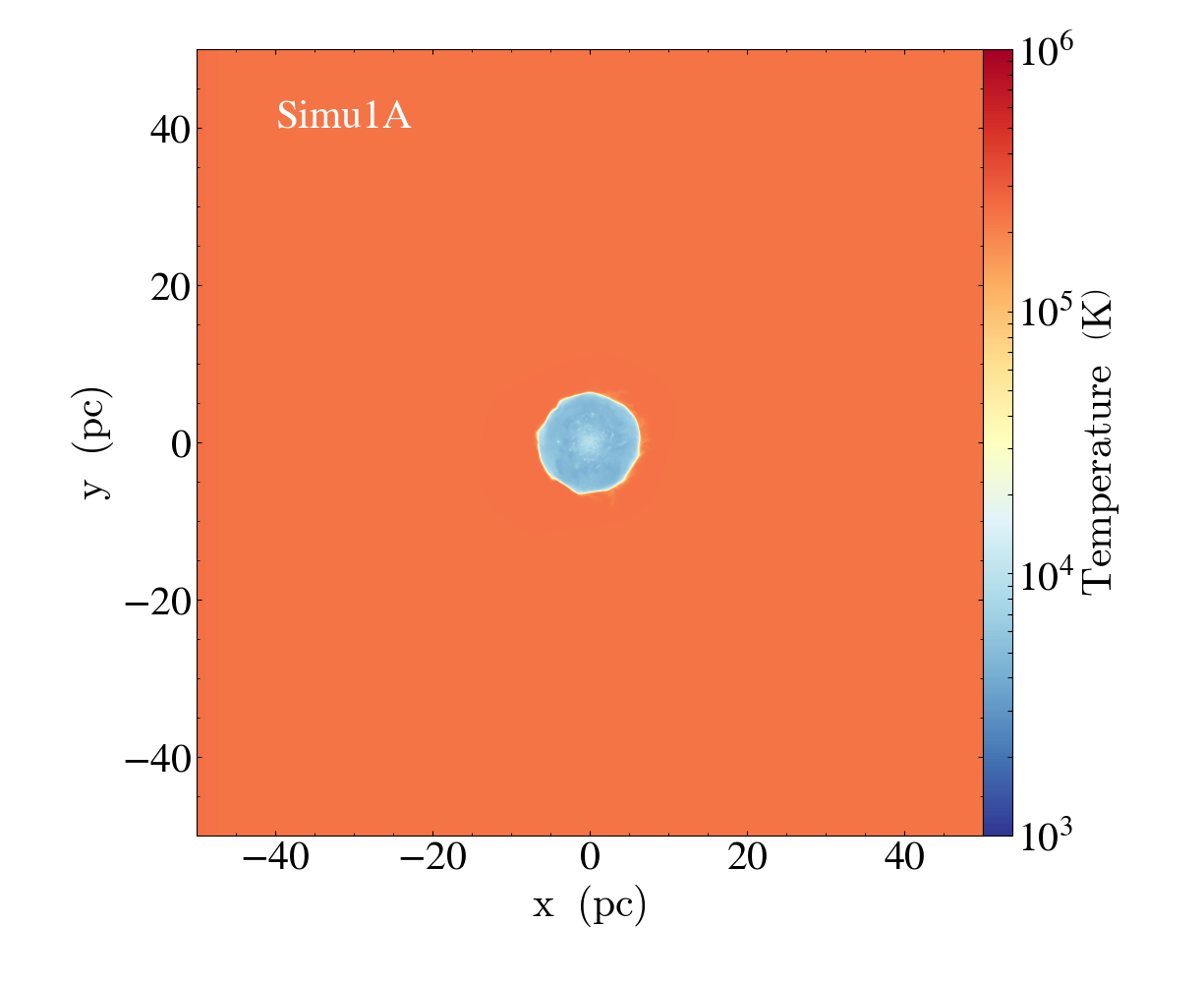} \\      
    \includegraphics[width=0.49\textwidth]{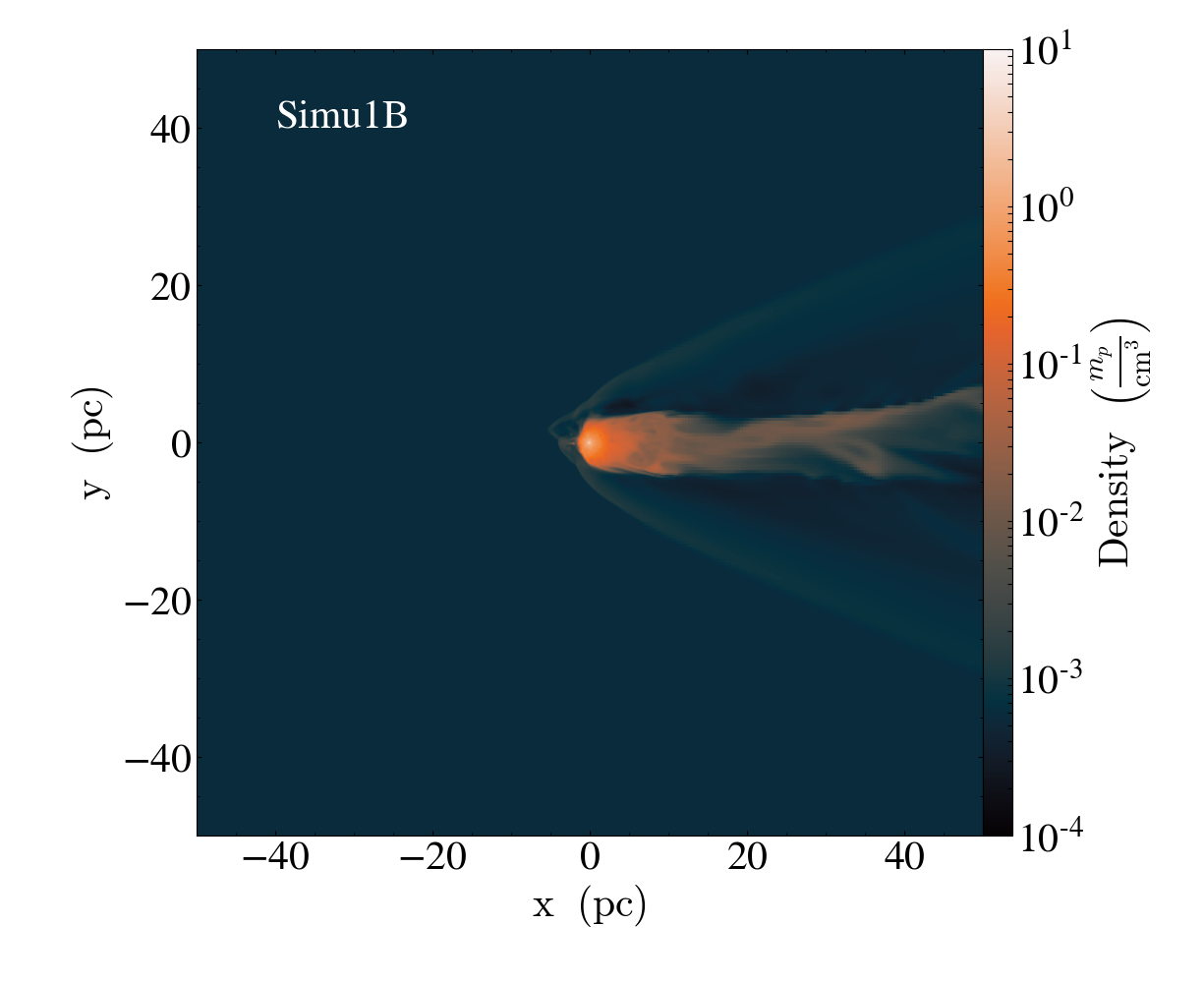}  
    \includegraphics[width=0.49\textwidth]{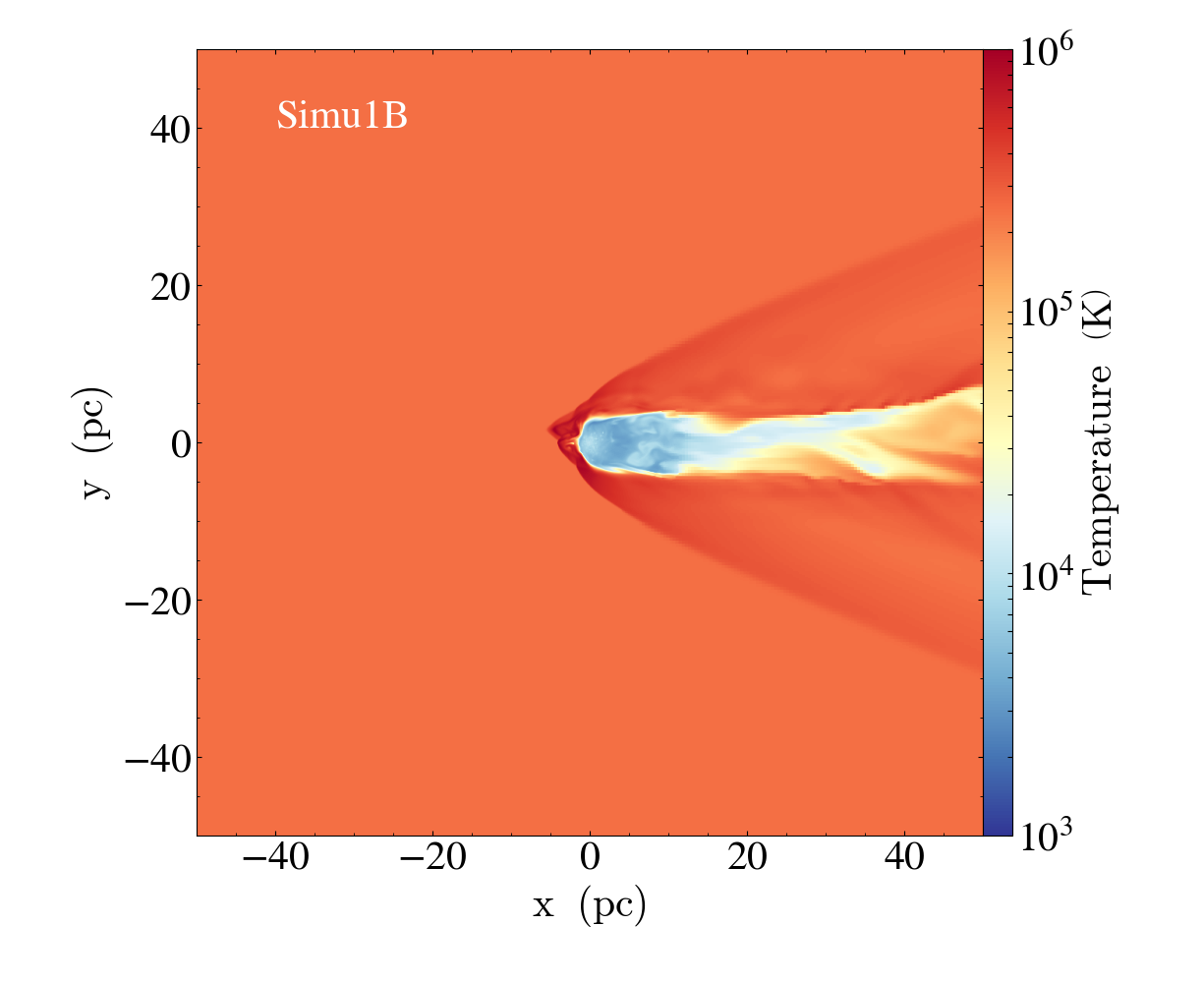} \\  
    \includegraphics[width=0.49\textwidth]{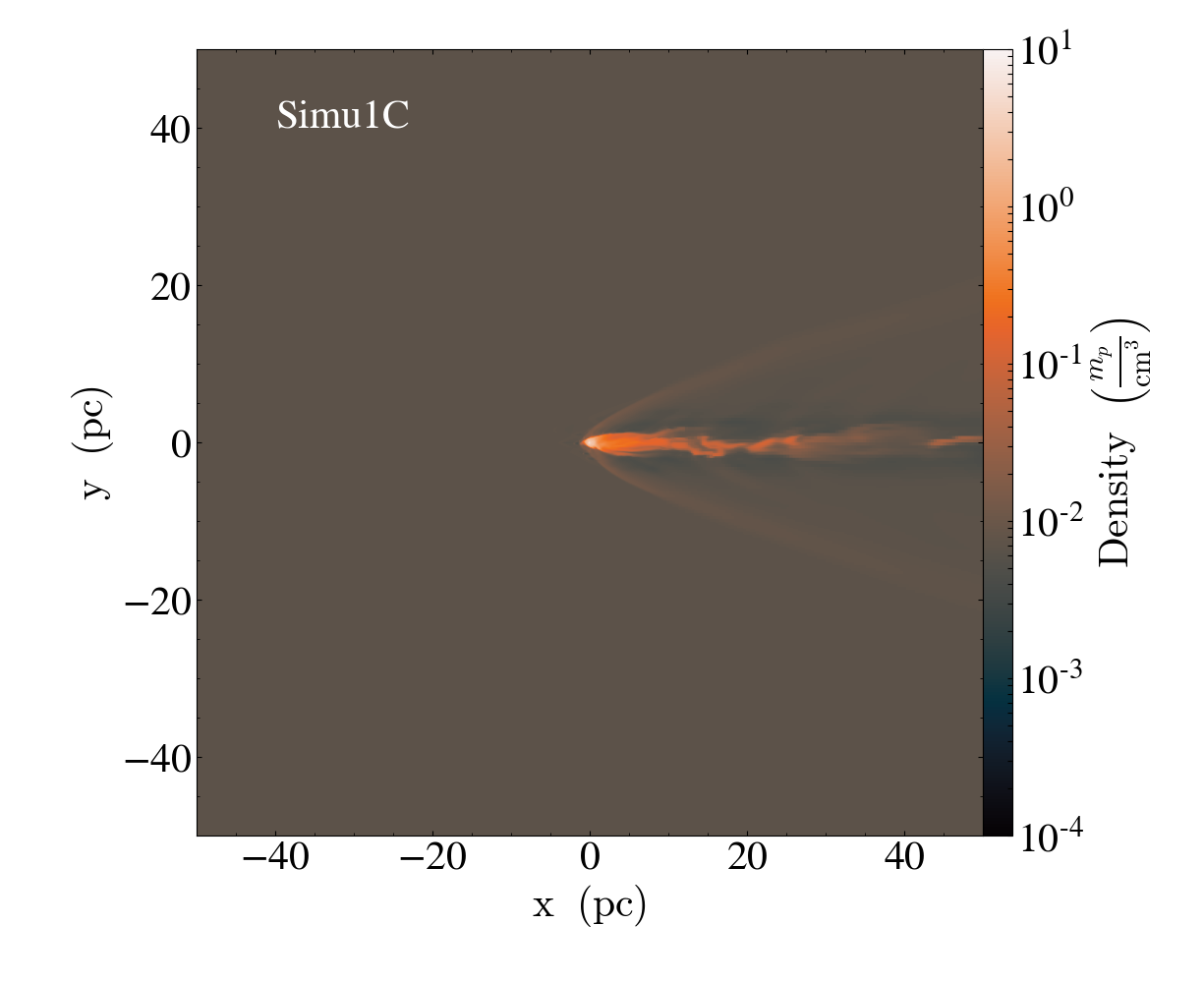}  
    \includegraphics[width=0.49\textwidth]{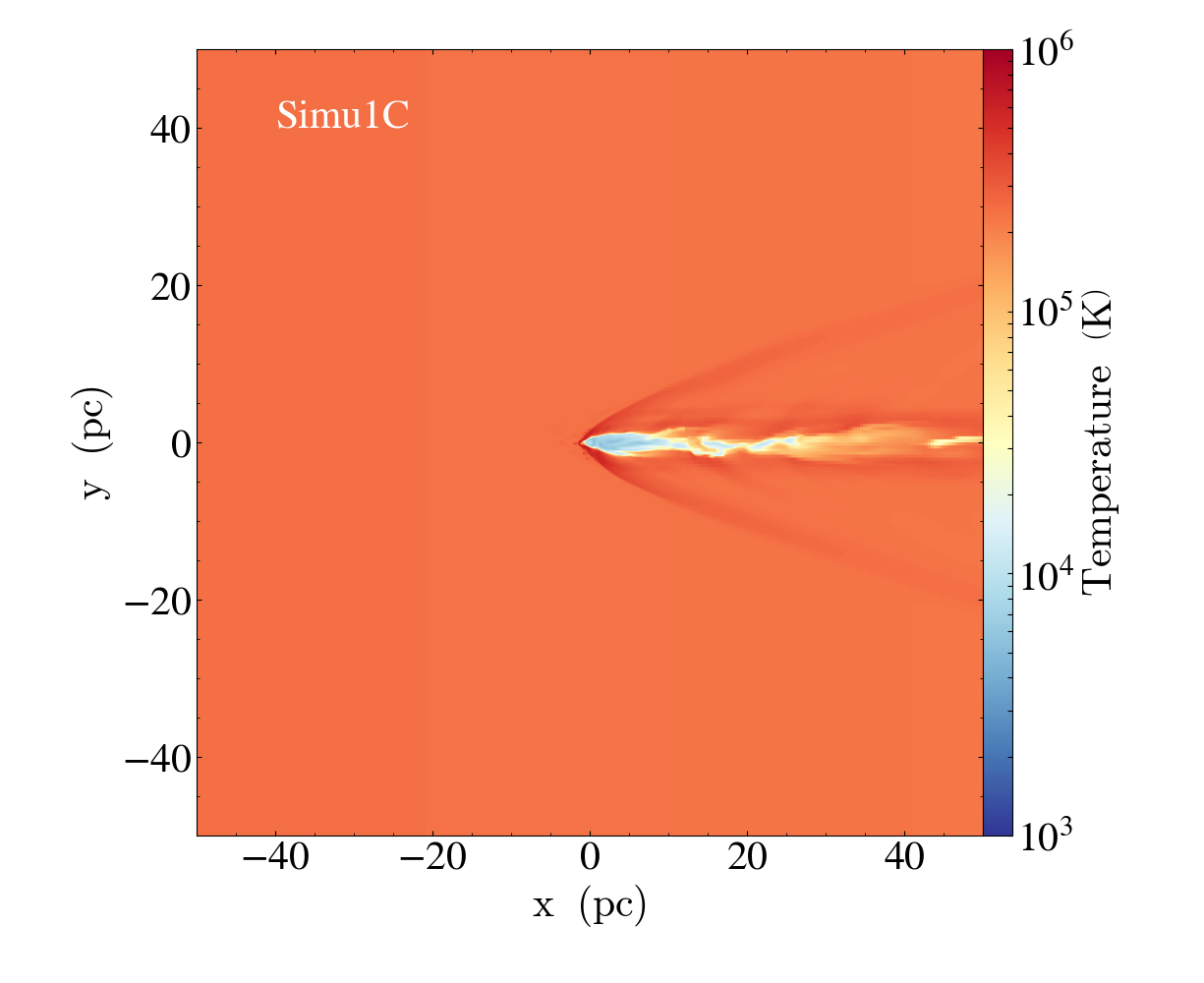}    
    \caption{Snapshots at 5~Myr displaying the density (left) and temperature (right) maps of simulations \emph{Simu1A}, \emph{Simu1B} and \emph{Simu1C} (from top to bottom, respectively).}
    \label{Figure:maps}
\end{figure*}

\cite{Priestley11} showed that the motion of typical intermediate-mass GCs through the Galactic halo medium reduces the ICM content thanks to ram-pressure stripping. Thus we discuss a first set of models to test the effect of the different parameters and processes, especially the ram-pressure stripping mechanism, on the ICM evolution in typical intermediate-mass GCs. These parameters are the velocity of the GC, the halo medium density and the temperature of the stellar winds.\footnote{We did not investigate the effect of the stellar mass-loss rate parameter in this study and used canonical values which are favoured in the literature. However, it can have an effect on the results \citep[as discussed in][]{Priestley11}.} We list in \autoref{table:simulations} the different simulations presented in this paper with their main characteristics. More details are presented in the following sections. We run each simulation until the mass of the ICM within 5~pc over the total stellar mass-loss (cf. \autoref{Figure:mass}, right panel) reaches a stable behaviour (e.g. gas accumulation) or an asymptotic behaviour indicating a steady state. We choose this radius as it is a good approximation for the half-mass radius of globular clusters in general \citep{Baumgardt18,deBoer19}. Indeed, most of the observations to detect the ICM in GCs are focused on their core region and are done within this radius.

\subsubsection*{Simu1A: control simulation}

We use \emph{Simu1A} as a control simulation for this set, we include hot stellar winds in a GC at rest in the halo where no ionising source is present (cf. \autoref{table:simulations}). \\

We show the mass accumulated within 5~pc as a function of time for our different simulations of the first set in \autoref{Figure:mass}\footnote{It is worth noting that all the mass injected through stellar winds is done within the inner 5~pc of the cluster.}. After $\sim$5~Myr, most of the ICM gas cools and sinks to the centre of the cluster (\autoref{Figure:mass}), which means that the energy from the stellar winds is not sufficient to heat and expand the gas beyond the half-mass radius (see \autoref{Figure:maps}). To summarise, in this simulation nearly all the ICM injected into the GC by stellar winds is retained, including a significant amount in the cluster centre (we choose for every simulation an arbitrary radius of 2.5~pc for the central region). We summarise the ICM mass in this central area and within 5~pc at 5~Myr for all the simulations in \autoref{table:core_properties}.

\subsubsection*{Simu1B: with ram-pressure stripping}

From now on, in all the simulations, the stellar cluster is at rest in the centre of the simulation volume (GC's frame of reference) and we introduce a continuous flow of the halo medium in the x-direction. We use an orbital velocity of the GC of 200 km~s$^{-1}$. To do that we assume that the gas in each cell of the simulation has a velocity of v$_\mathrm{x}$ = 200 km~s$^{-1}$ at t=0. The other parameters are identical to the ones used in \emph{Simu1A}, i.e. we include hot stellar winds and no ionising source is present. \\

We see in \autoref{Figure:maps} that the GC is stripped in its central region ($\sim$3~pc). The motion of the GC through the ambient environment creates a bow shock with a temperature at $\sim10^6$~K. If we now take a look at \autoref{Figure:mass}, right panel, the flattening of the curve means that there is equilibrium between the ram-pressure and the gravity forces. Despite a large part of the gas being stripped via the ram-pressure effect (more than half of the total injected gas already at 20~Myr), most of the ICM is located in the core region of 2.5~pc ($>96~\%$ of the mass), the rest is located in the tail of the GC, composed of the stripped material (\autoref{Figure:maps}). Therefore, even if the ram-pressure stripping allows to get rid of a large part of the ICM, most of the remaining gas cools and sinks in the central region of the cluster and one needs an additional mechanism to limit the amount of ICM to levels similar to what it is observed. 

\subsubsection*{Simu1C: denser halo}

In this simulation, we test if the ram-pressure can strip all the ICM by increasing the density of the ambient medium. We include hot stellar winds, no ionising source is present and the halo is denser by one order of magnitude ($\rho_{halo} = 6 \times 10^{-3}$ cm$^{-3}$). \\

We see in \autoref{Figure:maps}, that most of the gas is stripped away already at 5~Myr, but not all of it (down to a radius of $\sim$0.8~pc). Only 0.3~M$_\odot$ is retained within the inner 5~pc of the cluster and only $\sim$0.12~M$_\odot$ in the central region of the cluster (within 2.5~pc). This value is compatible with observational upper limits found in GCs often lower than 1~M$_\odot$ \citep[e.g.,][]{Roberts88,Smith90,vanLoon06,Boyer08,Matsunaga08,Barmby09,vanLoon09}.

We can conclude that the ram-pressure stripping due to the orbital motion of the GC in a rather dense halo medium is an efficient mechanism to limit the ICM in the central part of the stellar cluster at levels similar to observations.  

\subsubsection*{Simu1D: cool stellar winds}

The temperature of the winds we used so far is 6,000~K which corresponds to effective temperature of RGB stars in the metal-poor regime of Galactic GCs. The temperature of the wind and its velocity contribute to the energy of the cell in which the star resides. Thus it might have an effect on our result since it changes the resulting pressure. In this simulation we use a cooler wind temperature of 4,000~K, which is typical of the effective temperature of stars on the upper RGB close to the tip in metal rich globular clusters, to test the effect of this parameter. We do not include any ionising source and the GC is moving in a tenuous halo as in \emph{Simu1B}. \\

As expected, the decreased energy from the winds' lower temperature leads to similar results for the ICM values within the inner 5~pc but the mass accumulated is slightly larger than in \emph{Simu1B} (\autoref{Figure:mass}). We have tested here two rather extreme values of the temperature spanning a range of 2000~K and we conclude that the temperature of the winds has only a limited effect. 

\subsection{Massive GC - Simu2}\label{Simu2}

\begin{figure*}
    \centering
    \includegraphics[width=0.45\textwidth]{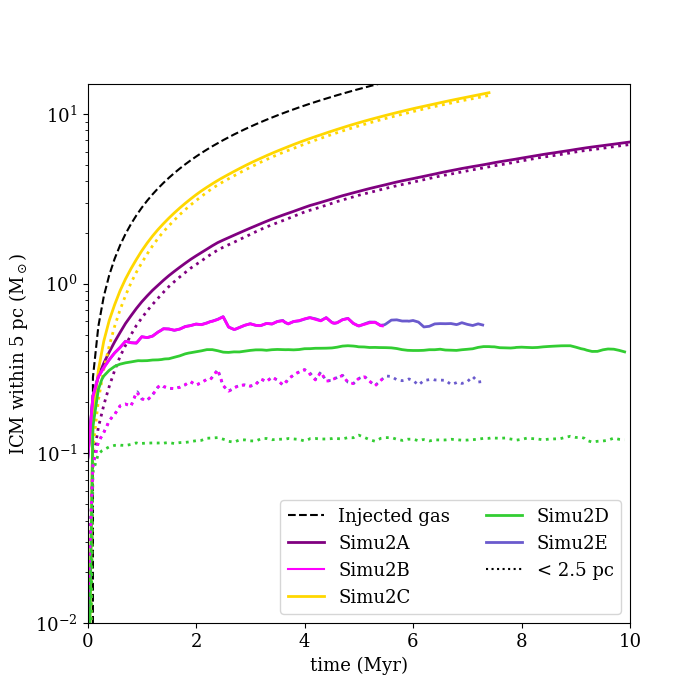}  
    \includegraphics[width=0.45\textwidth]{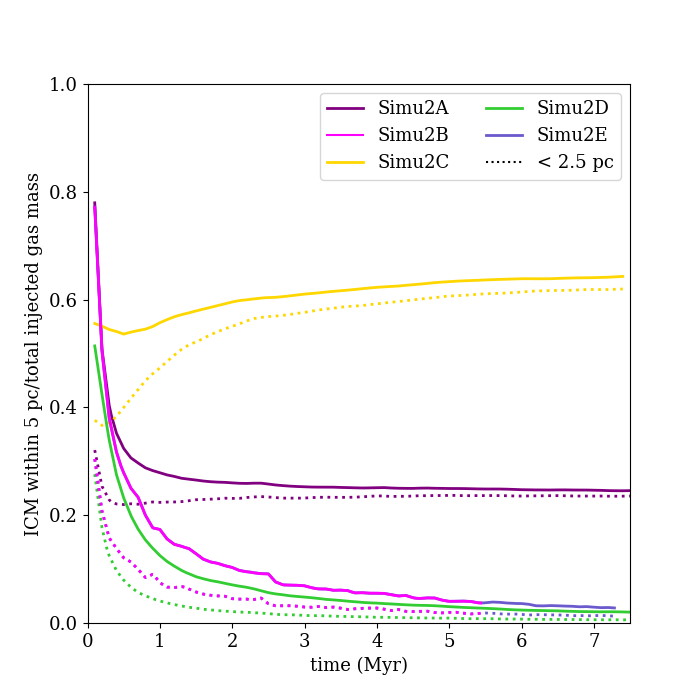} 
    \caption{\textit{Left:} ICM mass within 5~pc and the central part of 2.5~pc (solid and dotted lines, respectively) for the second set of simulations as a function of time. \textit{Right:} ratio of the ICM mass within 5~pc over the total mass injected by stellar winds for the second set of simulations as a function of time.}
    \label{Figure:mass2}
\end{figure*}

\begin{figure*}
    \centering
    \includegraphics[width=0.49\textwidth]{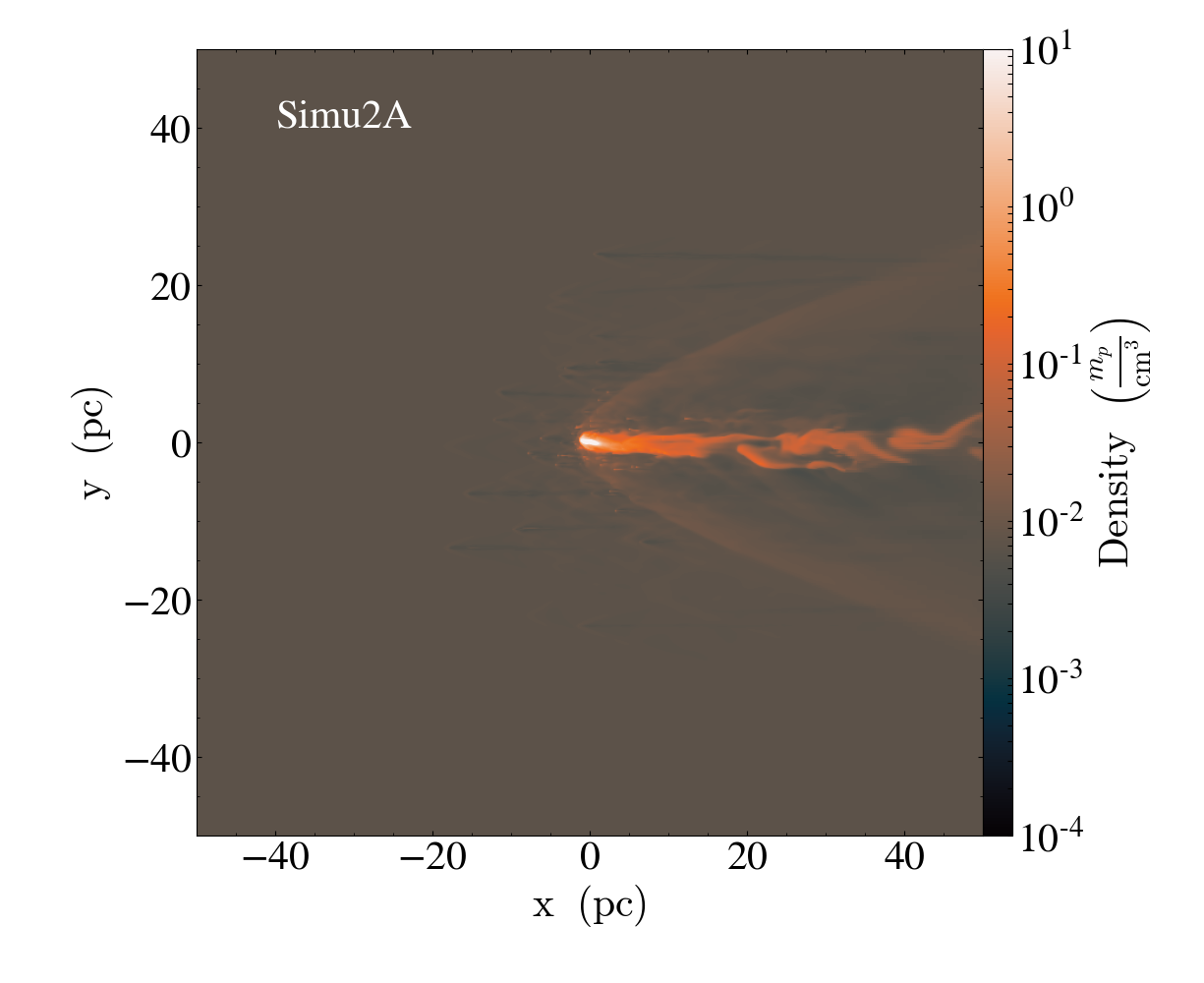} 
    \includegraphics[width=0.49\textwidth]{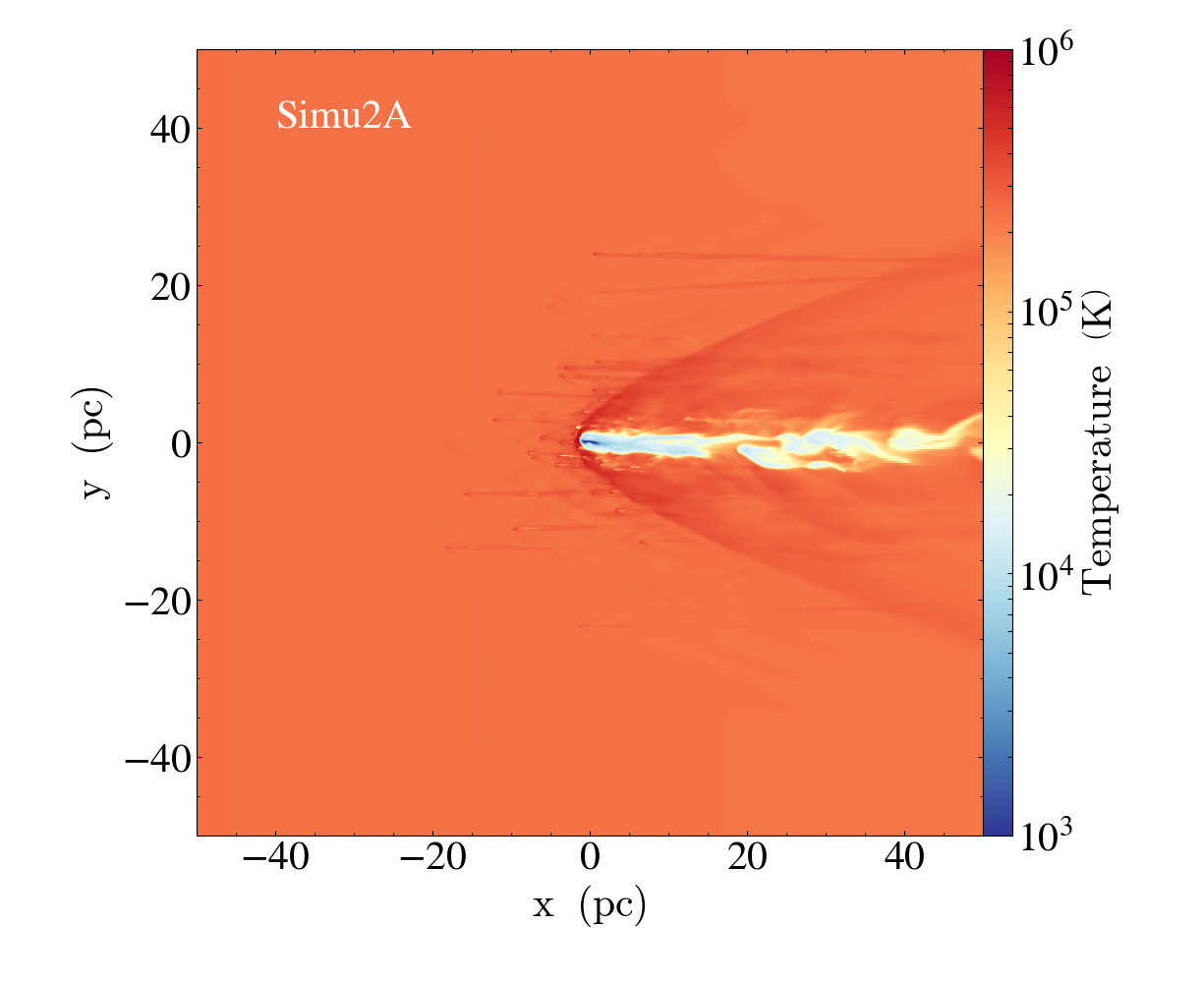} \\ 
    \includegraphics[width=0.49\textwidth]{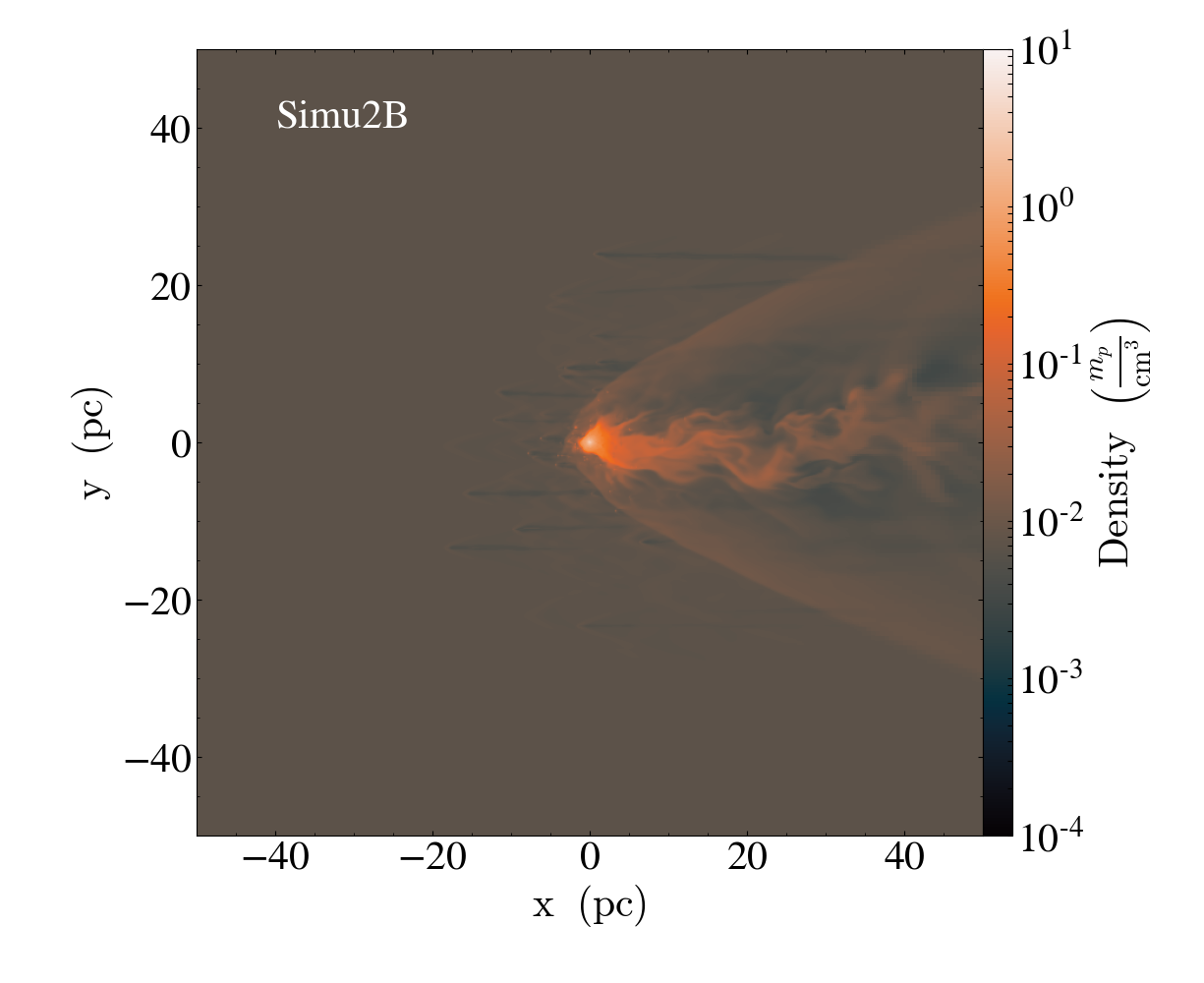} 
    \includegraphics[width=0.49\textwidth]{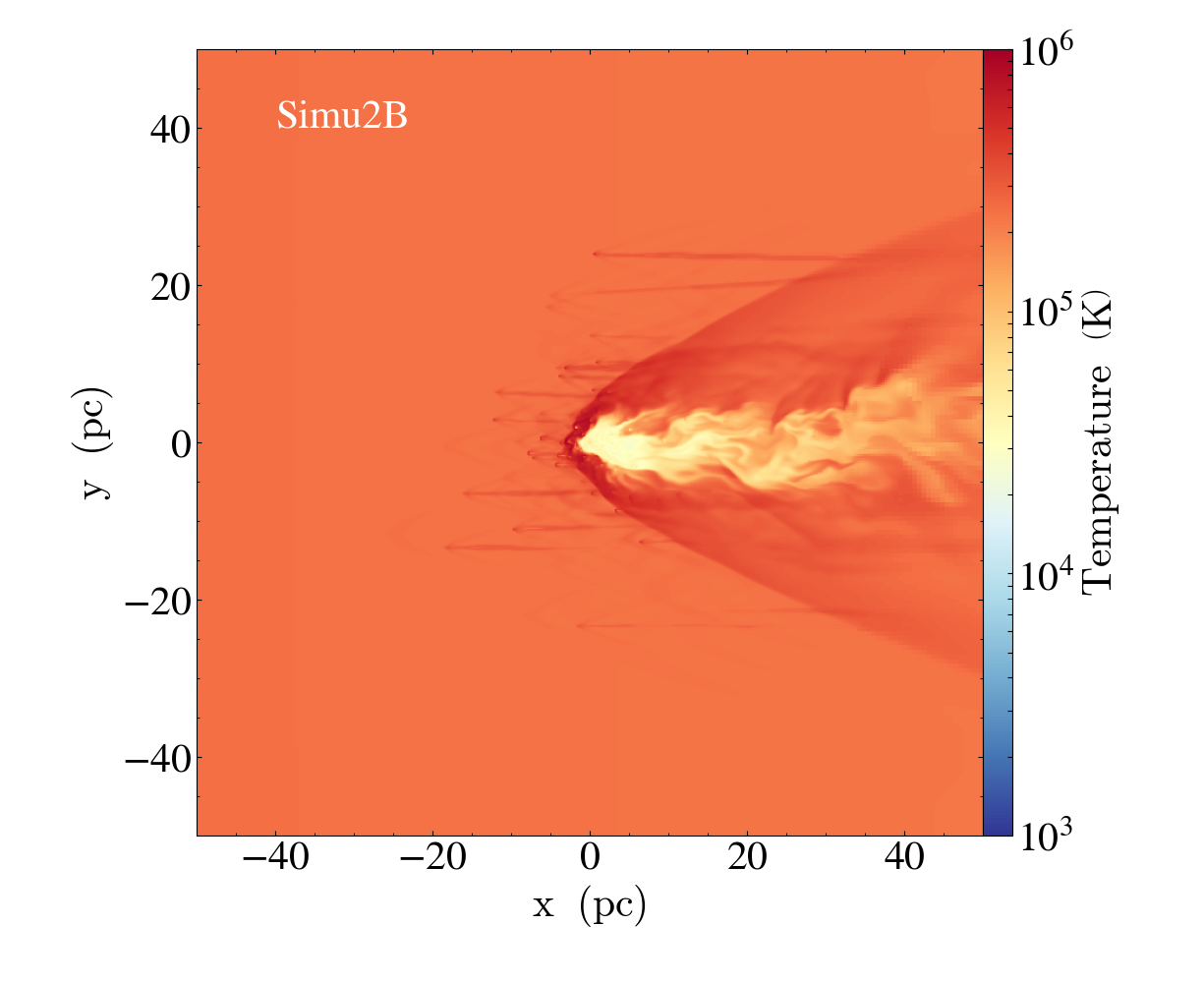} \\       
    \caption{Snapshots at 5~Myr displaying the density (left) and temperature (right) maps of simulations \emph{Simu2A}, \emph{Simu2B}, \emph{Simu2C} and \emph{Simu2D}.}
    \label{Figure:maps2}
\end{figure*}

\begin{figure*}
    \centering
    \includegraphics[width=0.49\textwidth]{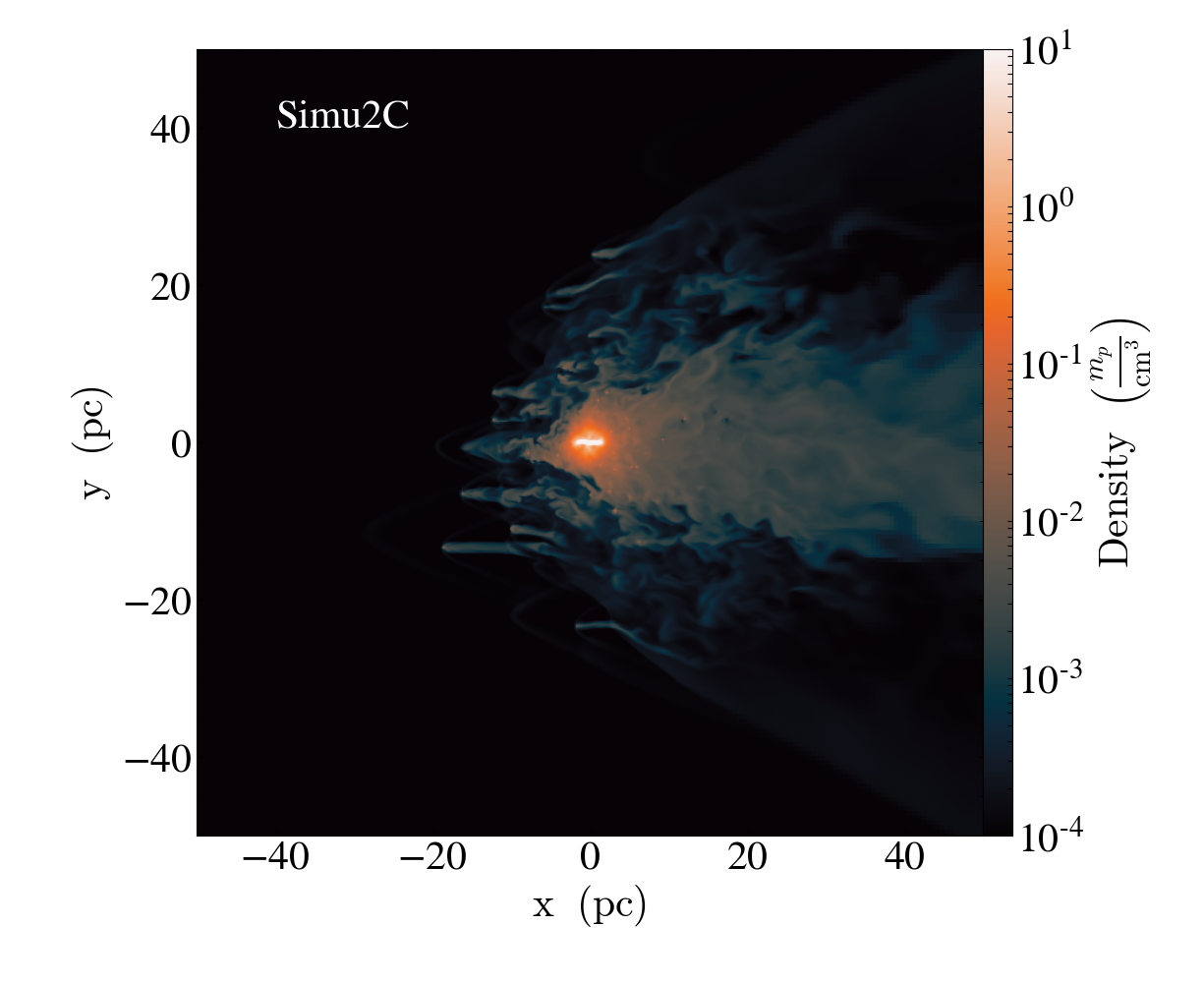} 
    \includegraphics[width=0.49\textwidth]{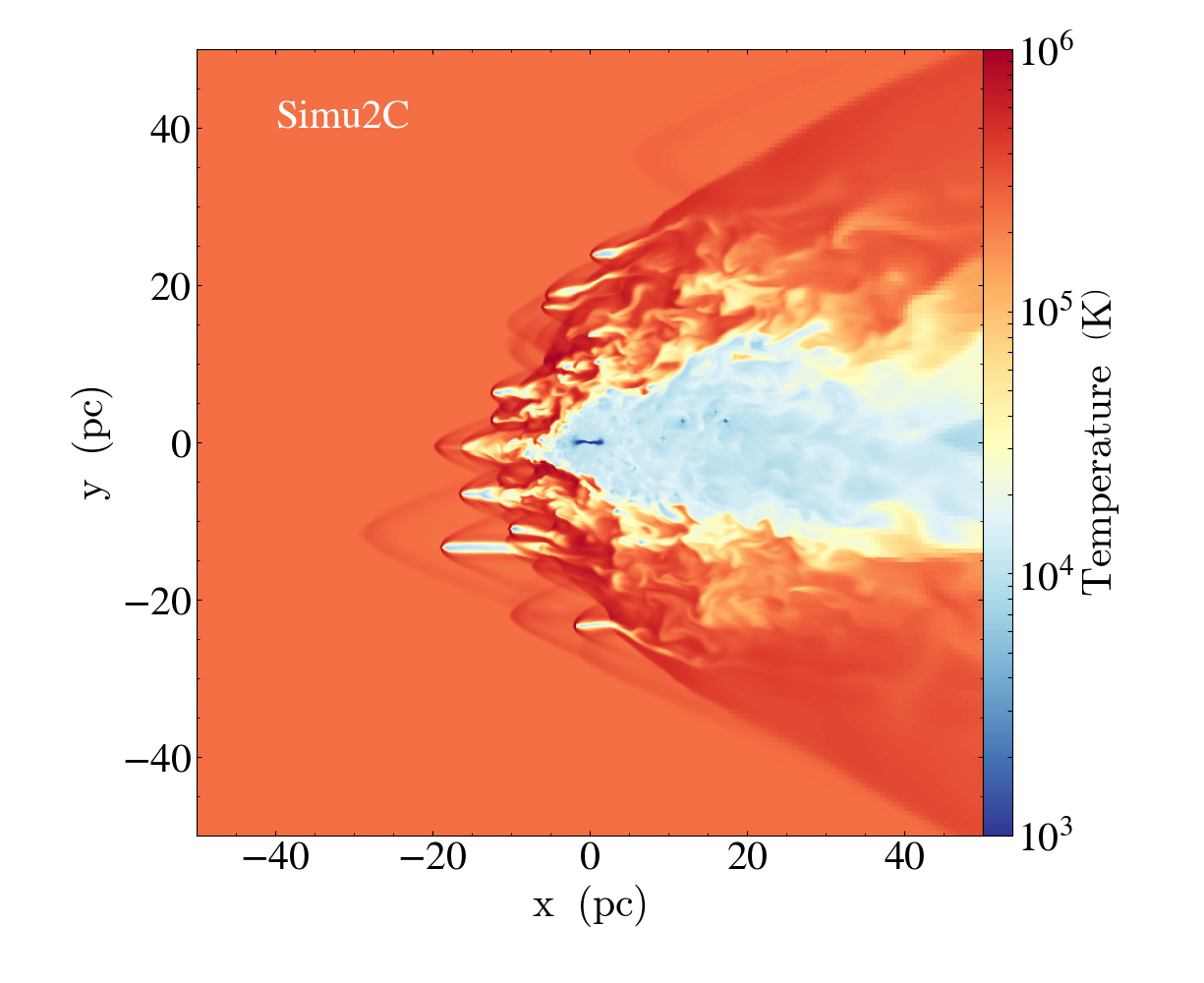} \\ 
    \includegraphics[width=0.49\textwidth]{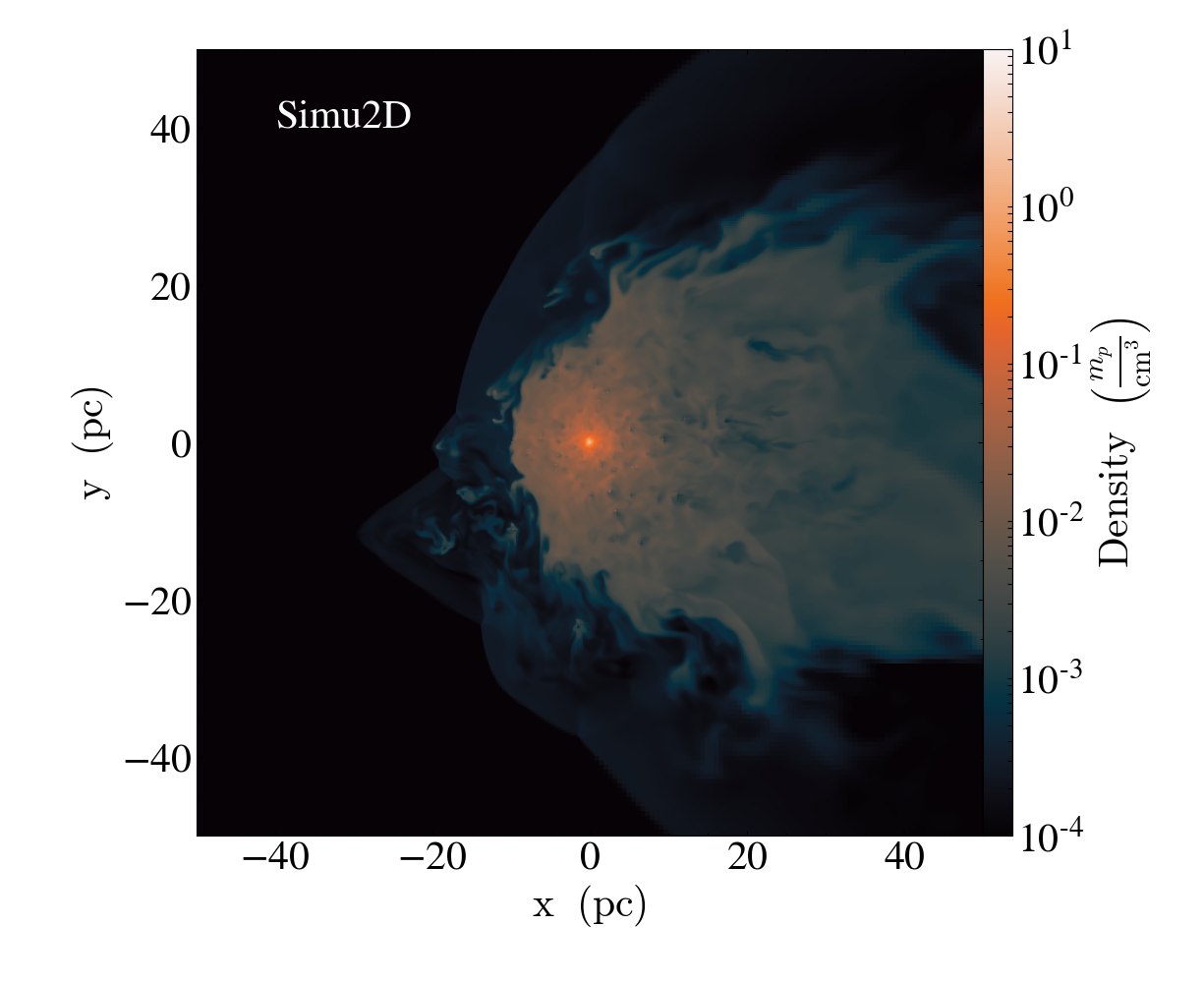} 
    \includegraphics[width=0.49\textwidth]{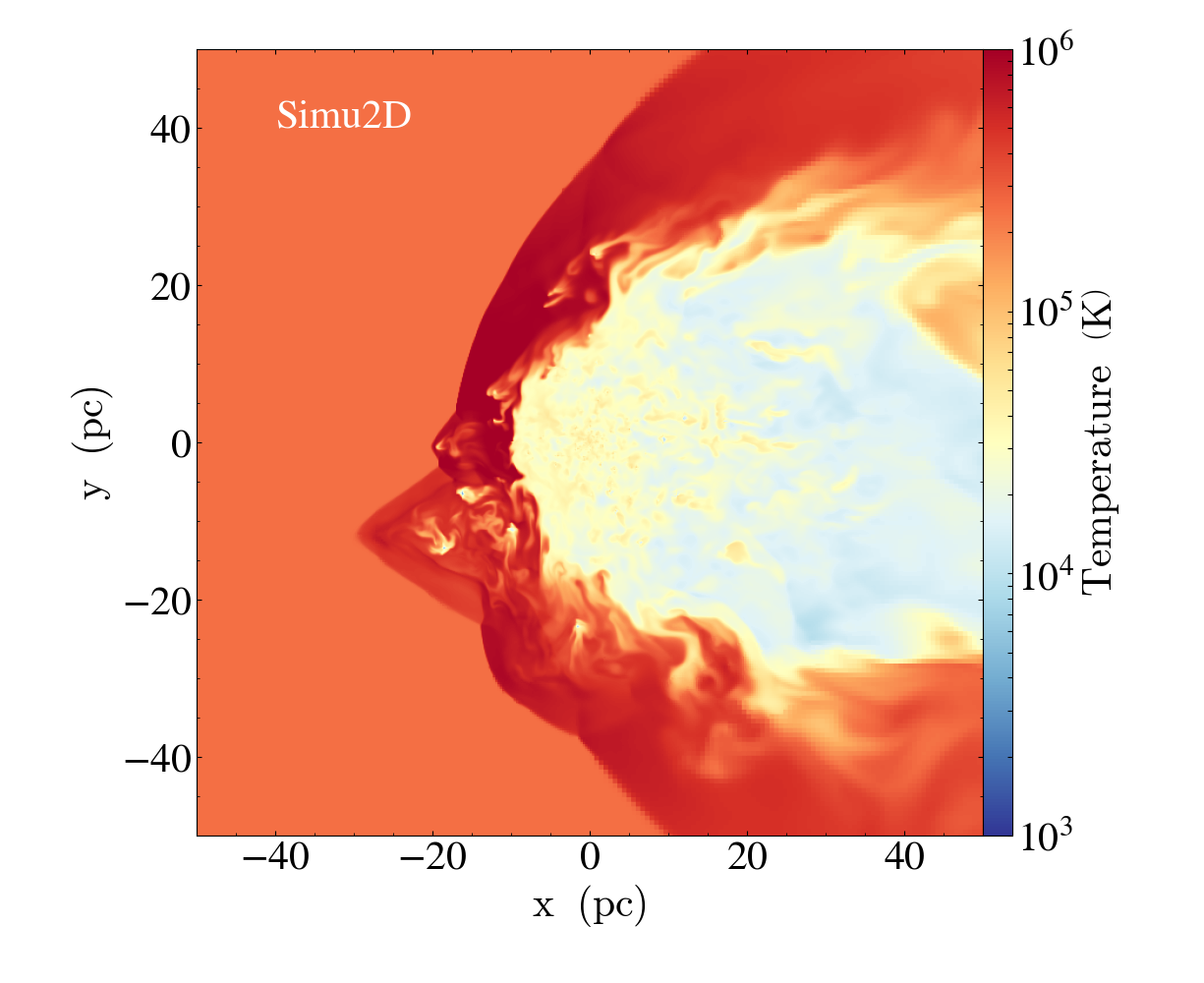} \\       
    \contcaption{}
\end{figure*}

We have shown that for typical intermediate-mass GCs a dense medium is necessary to maintain the ICM down to levels similar to observational limits. Now we look at massive GCs and we specifically focus on 47~Tuc \citep[$1.1\times10^6$~M$_\odot$,][]{McDonald15}. We expect that it will be more difficult to limit the ICM in massive GCs since the stellar mass-loss is higher and they have a deeper gravitational potential. 

We run several simulations to test the effect of different parameters on the gas evolution within massive stellar cluster, these parameters are the UV flux from ionising sources and the halo medium density.

\subsubsection*{Simu2A: control simulation}

In this simulation, as in the previous set we include hot stellar winds and no ionising source is present. However, the halo density is higher (7$\times 10^{-3}$~cm$^{-3}$) in agreement with the properties of the Galactic halo medium where the globular cluster 47~Tuc evolves \citep{Taylor93,McDonald15}. This density is similar to what is used in \emph{Simu1C} and was the key parameter to efficiently strip through ram-pressure most of the stellar cluster's ICM in a typical intermediate-mass GC to levels similar to observations. We use \emph{Simu2A} as a control simulation for this set. \\

The gas from the stellar winds\footnote{It is worth noting that most of the mass injected through stellar winds is done within the inner 5~pc of the cluster.} quickly cools and sinks into the centre of the cluster. There is 6.6~M$_\odot$ of gas within 2.5~pc after only 10~Myr and 6.8~M$_\odot$ within 5~pc, meaning that most of the gas left is located in the very central region of the cluster (\autoref{Figure:mass2}). This is also seen in \autoref{Figure:maps2} (top panel) and \autoref{Figure:profile}, where we see that the GC is nearly entirely stripped, but the gas in the very centre is dense and cool. 

Thus even if ram-pressure stripping allows to get rid of a large part of the gas\footnote{47~Tuc crossed the Galactic disk around 30~Myr ago \citep{Gillett88}, thus we would expect the stellar winds to contribute to about 84~M$_\odot$ to build up the ICM within the inner 5~pc.} (more than 70~\% of the total injected gas, \autoref{Figure:mass2}, right panel), the remaining part builds up a non negligible ICM and is located in the very central region of the cluster.

\subsubsection*{Simu2B: UV ionising flux}

\cite{McDonald15} have shown that UV radiation from hot post-AGB stars and cooling white-dwarfs of 47~Tuc can keep its ICM ionised. In turn, this thermalised ICM will expand well beyond the half-mass radius. To test this we add a post-AGB star as ionising source in our simulation (see Sect.~\ref{ramses} for more details on the ionising source). We include hot stellar winds and consider a dense halo environment. \\

\begin{figure*}
    \centering
    \includegraphics[width=0.6\textwidth]{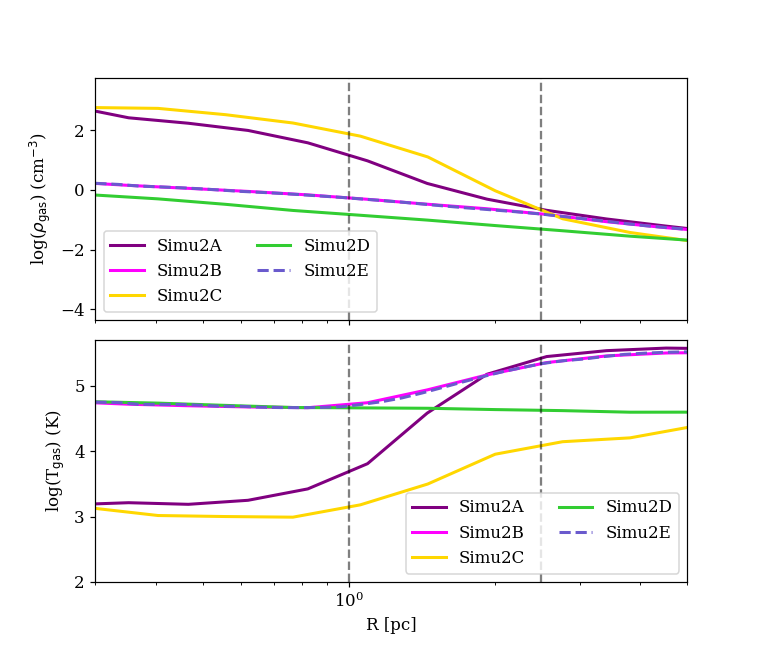} \\ 
    \caption{Density (top) and temperature profiles (bottom) for \emph{Simu2A}, \emph{Simu2B}, \emph{Simu2C}, \emph{Simu2D} and \emph{Simu2E} at 5~Myr. The dashed  vertical lines correspond to radii of 1~pc and 2.5~pc, respectively.}
    \label{Figure:profile}
\end{figure*}

As we can see in \autoref{Figure:maps2}, second panel, and in more details in \autoref{Figure:profile}, the density in the centre of the cluster at 5~Myr is lower, up to 2 orders of magnitude, compared to the case where there is no ionising source (\emph{Simu2A}). This is due to the radiative heating and pressure of the gas from the ionising source allowing the gas to expand quickly outwards (the ICM is completely ionised). In this case, the mass accumulated  within its 5~pc reaches quickly ($\sim$2~Myr) an asymptotic value of 0.6~M$_\odot$. In the centre of the cluster, i.e. 2.5~pc and 1~pc, this value is as low as $\sim$0.26~M$_\odot$ and $\sim$0.075~M$_\odot$, respectively. We expect about 11.2~$M_\odot$ of material lost by the stellar winds over 4~Myr within the inner 5~pc of the cluster. It means that 10.6~$M_\odot$ is located well beyond the half-mass radius of the cluster at 4~Myr ($\sim$95~\% of the total stellar mass-loss).

It confirms the results of \cite{McDonald15} who predicted that $\sim$11.3~$M_\odot$ of ICM within 47~Tuc should be cleared over $\sim$4~Myr. In addition it is in agreement with \cite{Freire01} and \cite{McDonald15} who claimed that all the gas left within 47~Tuc should be completely ionised. Finally, these results are  similar to what was found by \cite{Freire01} who reported a possible total ICM mass of $\sim$0.1~M$_\odot$ of gas within the inner 2.5~pc (0.26~M$_\odot$ in our case) and \cite{Abbate18} calculated a total mass of gas in the inner 1 pc of 47~Tuc of 0.023$\pm$0.005~M$_\odot$ (0.075~M$_\odot$ in our case). 

We compare our results with observations at $\sim$7~Myr in our simulation whereas 47~Tuc crossed the Galactic disk around 30~Myr ago \citep{Gillett88}. However, the ICM within the inner 5~pc and 2.5~pc reached a steady state (\autoref{Figure:mass2}, right panel) thus we do not expect differences between 7 and 30~Myr. 

\subsubsection*{Simu2C: tenuous halo}

We have shown in the case of typical intermediate-mass GCs that the halo density is a key parameter for the ICM ram-pressure stripping. We used in \emph{Simu2A} and \emph{Simu2B} a high halo density which corresponds to 47~Tuc's environment ($\sim$7.4~kpc from the Galactic centre). However, the halo density decreases quickly as a function of the distance from the Galactic centre. Thus here we aim to also test the effect of ram-pressure stripping in massive clusters evolving in a tenuous environment, i.e. that would orbit further away from the Galactic centre\footnote{We assume here circular orbits for GCs.}. It allows us to provide predictions for a broad range of stellar cluster's properties. 47~Tuc is a GC located at $\sim$7.4~kpc from the Galactic centre. If we assume a $\beta$-model as the gas density profile for the Galactic halo \citep{Miller13}, then a halo density of $\sim$2 orders of magnitude lower to the one we used for 47~Tuc (previous simulations \emph{Simu2A} and \emph{Simu2B}) corresponds to a distance from the Galactic centre of $\sim$64~kpc. We then choose in this simulation a density of $\sim$2 orders of magnitude lower than before for the halo (1$\times 10^{-4}$~cm$^{-3}$) to simulate a stellar cluster located at $\sim$64~kpc from the Galactic centre. We include hot stellar winds and no ionising source is present in this simulation.\\

As in \emph{Simu2A}, the gas quickly cools and most of it sinks into the centre of the cluster (\autoref{Figure:mass2}, \autoref{Figure:maps2} and  \autoref{Figure:profile}). The lower halo density compared to the one used in \emph{Simu2A} limits logically the ram-pressure mechanism and the gas retained within the inner 5~pc of the cluster is more important. Most of the stellar winds are retained to form a 12.6~M$_\odot$ ICM within the inner 5~pc at 7~Myr (+160~\% compared to \emph{Simu2A}).

\subsubsection*{Simu2D: tenuous halo and UV ionising flux}

In this simulation we test the effect of ram-pressure stripping and ionisation in massive clusters evolving in a tenuous environment. The parameters are identical to the ones used in \emph{Simu2C}, i.e. we include hot stellar winds and we choose a low density halo, however we also include an ionising source (same source as in \emph{Simu2B}). \\

As we can see in \autoref{Figure:mass2}, due to the very low initial gas density, there is less gas in the cluster centre compared to \emph{Simu2B}. There is an ICM mass of 0.26~M$_\odot$ and 0.13~M$_\odot$ in the central region in \emph{Simu2B} and \emph{Simu2D}, respectively. The ICM mass is also very low in the half-mass radius region at 7~Myr ($\sim$0.4~M$_\odot$). 

Most of the studies to determine the ICM mass focus on the cluster centre (1-2.5~pc), thus we find low values of ICM in the cluster centre (0.026~M$_\odot$ and 0.13~M$_\odot$ at 1~pc and 2.5~pc, respectively) similar to what it is observed. \\

In this simulation we have shown that massive clusters located far from the Galactic centre display similar ICM values in their core compared to GCs evolving close to the Galactic centre like 47~Tuc.

\subsubsection*{Simu2E: lower UV ionising flux}

\cite{McDonald15} proposed an ionising photon rate lower by one order of magnitude than the one we used in our simulation, their UV flux from model ($\lambda <$ 912 \AA) being 2.43$\times 10^{47}$ s$^{-1}$. To numerically model their work and test their prediction we run the simulation \emph{Simu2B} with an ionising source providing the same UV flux ($\lambda < 912 \AA$). We include hot stellar winds, consider a dense halo environment, and the photoionisation rate of our source is 2.43$\times 10^{47}$ s$^{-1}$. \\ 

As expected, the results are similar to the ones of \emph{Simu2B}, i.e. temperature and density profile, as well as the gas mass accumulated in the stellar cluster (\autoref{Figure:mass2}, \autoref{Figure:profile}). This is due to the fact that both ionising photons rates used here (2.43$\times 10^{47}$~s$^{-1}$ and 5.7$\times 10^{48}$~s$^{-1}$) are already a few orders of magnitude higher than the rate needed to fully ionise the ICM \citep[1.6$\times 10^{44}$ photons s$^{-1}$ with wavelengths $<$ 912 \AA,][]{McDonald15}. 

\begin{table}
\centering
\begin{tabular}{ c c c}
	\hline 
	Simulation & 2.5~pc (M$_\odot$) & 5~pc (M$_\odot$) \\ \hline
    M$_\mathrm{winds}$ & - & {1.58} \\
    Simu1A \textit{(control simulation)} & 0.37 & 1.14 \\
    Simu1B \textit{(with ram-pressure stripping)} & 0.34 & 0.55 \\
    Simu1C \textit{(denser halo)} & 0.12 & 0.26 \\
    Simu1D \textit{(cool stellar winds)} & 0.40 & 0.61 \\ \hline
    M$_\mathrm{winds}$ & - & 14.0 \\    
    Simu2A \textit{(control simulation)} & 3.30 & 3.49 \\
    Simu2B \textit{(UV ionising flux)} & 0.26 & 0.56 \\
    Simu2C \textit{(tenuous halo)} & 8.48 & 8.90 \\ 
    Simu2D \textit{(tenuous halo and UV ionising flux)} & 0.13 & 0.42 \\
    Simu2E \textit{(lower UV ionising flux)} & 0.26 & 0.56 \\
    \hline 
\end{tabular}
\caption[]{ICM mass (M$_\odot$) within the centre (2.5 pc) and the typical cluster's half-mass radius  (5~pc) of the different simulations at 5~Myr. M$_\mathrm{winds}$ is the total mass lost by stellar winds at 5~Myr.} \label{table:core_properties}
\end{table}

\section{Discussion}\label{discussion}

We investigate stellar clusters of 10$^5$~M$_\odot$ and 10$^6$~M$_\odot$, with and without ram pressure stripping as well as photoionisation. We find that for most cases, the ram-pressure stripping alone is not enough to effectively remove the ICM from the clusters due to the fact that the gas concentrates in the central regions of the clusters (due to cooling). However, when including photoionisation, the gas is heated and expands to fill the full volume of the cluster (i.e. not concentrate in the cluster centre). Due to this increased surface area, the ICM can then be more efficiently stripped by ram-pressure, this mechanism working on very short time-scales ($\lesssim3$~Myr). \\

Our simulations with ram-pressure and without photoionisation showed that globular clusters with masses around $10^5$~M$_\odot$ display in general a low ICM in their centre. For more massive GCs ($\sim10^6$~M$_\odot$), the ram-pressure mechanism alone is not efficient enough to strip the ICM even in a dense environment (i.e. GCs orbiting within $\sim$10~kpc from the Galactic centre). However, the addition of ionisation by UV sources provide a UV flux ($\lambda < 912 \AA$) able to fully ionise the cluster's ICM and favour its loss. \\

Most of typical intermediate-mass GCs should also host a source with a photon rate high enough to ionise their ICM, then favouring its loss (similar to what happens in massive GCs to a lesser extent). Thus, if we take into account ram-pressure stripping and ionisation, we predict a negligible ICM within these typical  GCs and the values reported in Table~\ref{table:core_properties} (taking into account only ram-pressure stripping) are upper limits.

On the other hand, while taking into account ram-pressure stripping and ionisation, massive GCs will display a low amount of ionised ICM in their central region, in agreement with observations. \\

To summarise, we do not expect to find any significant ionised ICM in the core of globular clusters, regardless of their mass and their distance to the Galactic centre. However, this very low amount of ionised ICM should be detectable inside the core of massive GCs (M$\gtrsim 5 \times 10^5 $~M$_\odot$). These predictions can be tested by investigating the total ICM withing the inner 5~pc of the most massive GCs, such as 47~Tuc, NGC~2419, NGC~2808, NGC~5139, NGC~5824, NGC~6266, NGC~6273, NGC~6284, Liller~1, NGC~6388, NGC~6402, Terzan~5, NGC~6441, NGC~6715, NGC~6864 and NGC~7089.\\

The same mechanisms presented in this work could also be efficient at younger ages. Therefore, in a future work we will extend our simulations to study the early phases of young massive clusters and proto-GCs.

We have shown that without photoionisation, the gas cools and sinks to the centre, which minimises its cross-sectional area, and in turn, minimises the effect of ram-pressure. This result has  strong implications on scenarios trying to explain the origin and formation of multiple stellar populations within GCs. Indeed, some of these scenarios require that the gas in stellar clusters is retained for a non negligible period of time and be allowed to cool to form these different stellar generations \citep[e.g.,][]{DErcole08}. For instance in the AGB scenario, the second stellar population starts to form after the explosion of all type II SNe of the first stellar population at about 40~Myr and it lasts until $\sim$100~Myr. 

During the early evolution of the progenitor of globular clusters, the properties of the environment are expected to be different. In addition the stellar population of these very young clusters will provide a strong UV flux. Thus it is vital while investigating the early evolution of the ICM in GC progenitors and the formation of multiple populations to take into account the radiative feedback effects of stars \citep[see also][]{Conroy11}. For instance, if all the ionising sources of the stellar cluster are not taken into account and the full radiative transfer calculations are not included in the simulations on the formation of multiple populations, it might greatly favour the formation of the second stellar population in the centre of the cluster thanks to the cooling and sinking gas \citep{DErcole16,Calura19}.

\cite{Bastian14}, \cite{Hollyhead15}, \cite{Cabrera15} and \cite{Hannon19} have found that young massive clusters (YMC) were gas free within the first 2-4~Myr of their lives. Thus these stellar clusters are very efficient at clearing out their gas, at odds with theoretical studies without invoking special circumstances \citep[e.g., IMF variations, high star formation efficiency, strong coupling of the energy produced by stellar feedback to the gas,][]{Krause16}. The results of this study suggest that the combined effect of radiative heating and radiative pressure (from ionising sources) with ram-pressure stripping could be the processes at work in YMCs which would explain why they are gas free. \\

\section*{Acknowledgements}
W.C. acknowledges funding from the Swiss National Science Foundation under grant P400P2\_183846. P.B. acknowledges ERC starting grant 638707. N.B. and W.C. gratefully acknowledge financial support from the European Research Council (ERC-CoG-646928, Multi-Pop). N.B. gratefully acknowledges financial support
from the Royal Society (University Research Fellowship) We would like to thank Iain Mcdonald, Joki Rosdahl and Maxime Trebitsch for useful discussions. This research used: The Cambridge Service for Data Driven Discovery (CSD3), part of which is operated by the University of Cambridge Research Computing on behalf of the STFC DiRAC HPC Facility (www.dirac.ac.uk). The DiRAC component of CSD3 was funded by BEIS capital funding via STFC capital grants ST/P002307/1 and ST/R002452/1 and STFC operations grant ST/R00689X/1. The DiRAC@Durham facility managed by the Institute for Computational Cosmology on behalf of DiRAC. The equipment was funded by BEIS capital funding via STFC capital grants ST/P002293/1 and ST/R002371/1, Durham University and STFC operations grant ST/R000832/1. DiRAC is part of the National eInfrastructure. Finally, we would like to thank the referee for the pertinent questions, and suggestions that have greatly helped us improve the presentation of our results.

\bibliographystyle{mnras}
\bibliography{bib}

\end{document}